\documentclass[conference]{IEEEtran}
\IEEEoverridecommandlockouts

\usepackage{cite}
\usepackage{amsmath,amssymb,amsfonts}
\usepackage{algorithmic}
\usepackage{graphicx}
\usepackage{xcolor}

\usepackage[caption=false,font=footnotesize]{subfig}

\usepackage[hidelinks]{hyperref}

\begin{document}

\title{The Circulate and Recapture Dynamic of Fan Mobility in Agency-Affiliated VTuber Networks}

\author{\IEEEauthorblockN{Tomohiro Murakami}
\IEEEauthorblockA{\textit{Degree Programs in Business Sciences} \\
\textit{University of Tsukuba}\\
Tokyo, Japan \\
s2540416@u.tsukuba.ac.jp}
\and
\IEEEauthorblockN{Mitsuo Yoshida}
\IEEEauthorblockA{\textit{Institute of Business Sciences} \\
\textit{University of Tsukuba}\\
Tokyo, Japan \\
mitsuo@gssm.otsuka.tsukuba.ac.jp}
}

\maketitle

\begin{abstract}
VTuber agencies---multichannel networks (MCNs) that bundle Virtual YouTubers (VTubers) on YouTube---curate portfolios of channels and coordinate programming, cross appearances, and branding in the live-streaming VTuber ecosystem. It remains unclear whether affiliation binds fans to a single channel or instead encourages movement within a portfolio that buffers exit, and how these micro level dynamics relate to meso level audience overlap. This study examines how affiliation shapes short horizon viewer trajectories and the organization of audience overlap networks by contrasting agency affiliated and independent VTubers. Using a large, multiyear, fan centered panel of VTuber live stream engagement on YouTube, we construct monthly audience overlap between creators with a similarity measure that is robust to audience size asymmetries. At the micro level, we track retention, changes in the primary creator watched (\emph{oshi}), and inactivity; at the meso level, we compare structural properties of affiliation specific subgraphs and visualize viewer state transitions. The analysis identifies a pattern of \emph{loose mobility}: fans tend to remain active while reallocating attention within the same affiliation type, with limited leakage across affiliation type. Network results indicate convergence in global overlap while local neighborhoods within affiliated subgraphs remain persistently denser. Flow diagrams reveal \emph{circulate and recapture} dynamics that stabilize participation without relying on single channel lock in. We contribute a reusable measurement framework for VTuber live streaming that links micro level trajectories to meso level organization and informs research on creator labor, influencer marketing, and platform governance on video platforms. We do not claim causal effects; the observed regularities are consistent with proximity engineered by VTuber agencies and coordinated recapture.
\end{abstract}


\section{Introduction}

The creator economy, while promising meritocratic opportunity, operates as a volatile market for attention that creates systemic precarity for workers. Within this environment, talent agencies (often called multichannel networks, MCNs; hereafter, agencies) function not only as gatekeepers but as organizational infrastructures that buffer risk and stabilize careers at the level of a portfolio. They curate groups of creators, coordinate cross promotion, and bundle talent into constellations that are close in algorithmic space. Although this infrastructural view is conceptually plausible, it lacks robust empirical validation. Prior work often relies on indirect proxies for audience behavior or treats affiliation as a background covariate, which leaves a critical gap in understanding how meso level infrastructures shape fan trajectories and the topology of audience overlap over time \cite{Vrontis2021InfluencerReview, Gardner2016}.

To address these gaps, we assemble a large scale longitudinal dataset of deduplicated YouTube live chat logs and construct a fan centered panel that tracks unique users as they move between channels month by month. This design enables direct measurement of short horizon retention, changes in the primary creator watched (\emph{oshi}), and inactivity as observed behaviors. We complement this with a dynamic audience overlap network in which creators are nodes and shared viewers induce weighted edges. Monthly networks allow us to examine how meso level structure evolves in tandem with micro level movements.

Guided by these aims, we pose two linked research questions:
\begin{description}
  \item[RQ1.] How do fan retention, switching of the primary creator watched, and inactivity differ between agency affiliated and independent creators?
  \item[RQ2.] How are these behavioral dynamics associated with structural properties of the monthly audience overlap network when comparing the affiliated and independent subgraphs?
\end{description}
RQ1 is answered using a micro level, fan centered panel in which each viewer is assigned a single oshi per month based on the channel that receives the largest share of their chat activity. RQ2 uses the same underlying chat logs but projects them into a meso level creator--creator network, where viewers are free to interact with multiple channels in the same month and edges reflect this multi-oshi activity. Together, these questions provide complementary views on the same underlying ``circulate and recapture'' dynamic, linking individual fan trajectories to the evolving structure of the audience overlap network.

We explicitly align each contribution with a concrete gap in prior work: (i) MCN scholarship has rarely linked affiliation to \emph{longitudinal, micro level} fan trajectories—we fill this gap with a three year, fan centered panel that directly observes retention, oshi switching, and inactivity at scale; (ii) audience overlap studies are often \emph{static} or affiliation agnostic—we model \emph{monthly} overlap networks and compare segment specific evolution in degree, density, and clustering; (iii) affiliation is commonly relegated to a \emph{control variable}—we recenter it as a meso level infrastructural condition and provide a measurement design that links micro behavior to meso organization.

Our contributions are threefold. First, we connect MCN research to the micro level trajectories of individual fans. Whereas prior work on creator agencies and MCNs has largely focused on creator attributes, case studies, or cross-sectional performance metrics, we construct a three-year fan-centered panel covering 458 VTuber channels and directly observe short-horizon retention, oshi switching, and inactivity at scale. Second, we extend audience-overlap network research by treating overlap as a dynamic structure shaped by affiliation rather than a static snapshot. While existing studies often analyze single-time audience networks or abstract away from organizational status, we build monthly overlap graphs and systematically compare degree, density, and clustering between agency and independent segments over time. Third, we recenter affiliation from a background control to a meso-level infrastructure variable and provide a transparent measurement framework that links micro level fan trajectories to meso level network organization. This design clarifies how MCN-like infrastructures modulate the “circulate and recapture” dynamics of fan mobility and offers a basis that subsequent causal analyses can build on.

\section{Literature Review}\label{sec:lit}

\subsection{VTubers and MCNs as Organizational Infrastructure}
The creator economy is increasingly shaped by organizational infrastructures that mediate relationships between creators and platforms \cite{Vrontis2021InfluencerReview, Joshi2025}. Virtual YouTubers (VTubers), a rapidly growing segment, exemplify this trend. Their success is not determined solely by content appeal or parasocial interaction; it is also strongly associated with corporate affiliation \cite{TanGreene2025}. Scholarship frames multichannel networks (MCNs) as ``cultural infrastructures'' that provide creators with technical, economic, and administrative leverage \cite{Gardner2016, Zhang2024MCN}. Concretely, MCNs are intermediary organizations that partner with multiple creators on a given platform and aggregate them into a managed portfolio, taking over business and marketing functions such as audience development, cross-promotion, brand deals, rights management, and legal or technical assistance in exchange for a share of the revenue \cite{Gardner2016, Zhang2024MCN}. In the VTuber context, agencies and similar management firms fit this organizational definition, and we therefore use the terms ``agency'' and ``MCN'' interchangeably in what follows. This support structure can generate substantial advantages, including reputation spillover that is associated with improved performance among affiliated talent \cite{Kokkodis2023}. However, affiliation can also impose costs. It may intensify precarity by amplifying the ``output imperative'' \cite{GregersenOrmen2023}, constrain autonomy through algorithmic opacity \cite{HodlMyrach2023}, and deepen reliance on emotionally taxing relational labor \cite{Glatt2024, Takano2024}. This creates a tension between the stability that MCNs offer and the precarity that they may reinforce, a dynamic shaped by external shocks \cite{Wollborn2023} and platform incentives \cite{Yang2022}.

\subsection{Fan behavior and oshi loyalty}
Recent work on live streaming has begun to formalize ``oshi change'' in terms of exploration and exploitation dynamics. Matsui et al.\ analyze a large-scale Japanese live streaming platform and model users' balance between exploring new streamers and repeatedly watching familiar ones via a turnover-rate metric, showing a prolonged ``golden period'' of exploration before behavior stabilizes \cite{Matsui2025}. They also operationalize ``oshi'', a core supported streamer, using platform-defined thresholds over behavioral signals such as watch time, monetary support, and comments, and link shifts in turnover rate to the acquisition and replacement of oshi over time. While their setting is a general-purpose live streaming service rather than VTubers or MCNs, we draw on this study as a representative example of how exploration, exploitation behavior and oshi change can be formalized in behavioral log data, rather than as a direct empirical benchmark. In our analysis below, we adopt a simpler operationalization tailored to VTubers, treating each viewer's oshi in a given month as the channel to which they direct the largest share of their chat activity.

\subsection{Network Mechanisms and Coordinated Attention}
A primary mechanism through which MCNs exert influence is the strategic management of audience networks. Online social systems are structured by homophily (the tendency for individuals to connect with similar others) \cite{Khanam2023}. In creator ecosystems this pattern appears as audience overlap between channels with similar content or appeal \cite{Rejeb2022, Conde2023, Sokolova2020}. MCNs do not leave this to chance; they actively orchestrate proximity by intensifying collaborations among creators within their portfolios \cite{Koch2018}. This managed coordination is theorized to create dense clusters of fans and to facilitate an ``attention spillover'' effect that channels viewers among affiliated creators. The process can internalize network externalities and convert them into more stable and scalable audience flows. Our study examines this phenomenon by modeling the chat user overlap graph and by using metrics such as the Simpson coefficient and mean degree to test whether MCN affiliation is associated with a denser and more clustered local network structure.

\subsection{Fan Monetization and the Research Gap}
The viability of the MCN model depends on sustaining fans’ financial commitment, which has shifted from transactional exchanges toward ``patron like'' relationships driven by intrinsic and social motivations \cite{Volkmer2024, Kunigita2023}. This patronage is shaped by reciprocity \cite{Lee2019}, social proof \cite{Lu2021}, and the duality between dedication and constraint \cite{Chou2023}, with streamer communication style a key predictor of support \cite{Giertz2022}. The long term sustainability of this model remains in question, since longitudinal studies show that financial contributions can decline substantially over a viewer’s tenure \cite{Ma2022}. To date, research has examined MCN infrastructure, network effects, and monetization psychology largely in isolation. A critical gap remains in understanding how these phenomena interconnect. This study addresses the gap by integrating multimodal engagement data (chats, donations, memberships) with dynamic network analysis over a three year panel that covers 458 channels from 2022 to 2025. By triangulating behavioral trajectories with evolving network structures, we provide an integrated empirical test of the view that MCNs function as infrastructures that stabilize attention and buffer market risk at the portfolio level.

\section{Data and Methodology}\label{sec:data_method}
\subsection{Data}

To examine how corporate infrastructure reshapes VTuber performance, we combine three publicly accessible data sources: (1) a community-edited roster from the \textit{FANDOM Virtual YouTuber Wiki}\footnote{\url{https://virtualyoutuber.fandom.com/wiki/Virtual_YouTuber_Wiki}}, (2) the official \textit{YouTube Data API v3}\footnote{\url{https://developers.google.com/youtube/v3}} for channel statistics and video metadata, and (3) archived live chat logs on YouTube retrieved via the command line utility \textit{yt\mbox{-}dlp}\footnote{\url{https://github.com/yt-dlp/yt-dlp}}. A seed roster of creators is collected from the community-edited \textit{FANDOM Virtual YouTuber Wiki}, whose list and profile pages provide channel identifiers, debut dates, avatar gender labels, and affiliation tags. As a community-maintained resource rather than an official registry, this wiki offers broad coverage of active VTubers but is unavoidably subject to omissions, outdated entries, and occasional tagging errors (for example, delayed updates when creators join or leave an agency). We therefore treat it as a high-recall seed list rather than a ground-truth label source and apply conservative filters (e.g., requiring valid YouTube channel IDs and debut dates within our sampling window). In cases where affiliation tags appeared ambiguous or conflicting, we either cross-checked channel self-descriptions on YouTube or excluded the channel from the analytical sample. These pages are scraped with a polite Python crawler that inserts delays of several seconds between requests. The roster is then supplied to the official \textit{YouTube Data API v3} to retrieve channel statistics and comprehensive video metadata, after which the command line utility \textit{yt\mbox{-}dlp} retrieves archived live chat logs that are not available through the API. All downloads were rate-limited and queued overnight to remain within the platform’s daily quota.

Data collection ran from November 2024 through February 2025 and targeted VTubers who debuted from January 2022 to December 2023. After excluding channels that were deleted or fully privatized by the end of the crawl, the corpus consists of a live chat panel including 458 creators and 55{,}476 archived streams. From these materials we recovered 190.3 million chat messages from 2.07 million distinct chatters, 790{,}139 Super Chat transactions from 103{,}259 patrons, and 7.20 million membership log entries for 339{,}308 unique subscribers (Table~\ref{tab:chat_data_overview}).

\begin{table}[tb]
  \centering
  \small
  \caption{Descriptive overview of the VTuber live-chat corpus.}
  \label{tab:chat_data_overview}

  \vspace{0.3em}
  \begin{tabular}{lr}
    \multicolumn{2}{l}{\textbf{(a) Basic counts}}\\
    \hline
    Metric & Full\\
    \hline
    Rows (chat--viewer pairs)   & 21,241,284\\
    Videos                      &     55,476\\
    VTubers                     &        458\\
    Total chats                 & 190,328,435\\
    Unique chat users           &   2,069,067\\
    \hline
  \end{tabular}

  \vspace{0.9em}
  \begin{tabular}{ll}
    \multicolumn{2}{l}{\textbf{(b) Data-collection window}}\\
    \hline
    First video date & 2022-01-08\\
    Last  video date & 2025-02-03\\
    \hline
  \end{tabular}

  \vspace{0.9em}
  \begin{tabular}{lccc}
    \multicolumn{4}{l}{\textbf{(c) VTuber counts by affiliation and gender}}\\
    \hline
                 & Female & Male & Total\\
    \hline
    Agency       & 263 &  78 & 341\\
    Independent  & 107 &  10 & 117\\
    \hline
    \textit{Total} & 370 &  88 & 458\\
    \hline
  \end{tabular}

  \vspace{0.6em}
  \parbox{0.95\linewidth}{\footnotesize}
\end{table}

Throughout the study we classify creators as either \emph{Agency affiliated} or \emph{Independent}. The latter label encompasses a wide spectrum of production models, ranging from hobbyists with minimal resources to small amateur studios and to former corporate talents who carry sizable existing followings (often termed ``super independents''). This heterogeneity inflates within group variance and may attenuate, or in rare cases exaggerate, the affiliation gaps reported below. Because reliable creator level data on prior experience, staffing, or external funding are not publicly available at scale, we refrain from further subclassification and analyze Independents as a single group. Future robustness checks could stratify Independents by observable proxies for organizational capacity (for example, debut day subscriber count or Twitter/X followers) or reestimate models after trimming extreme outliers. We leave these extensions for subsequent work and emphasize that the results presented here should be interpreted as \emph{average} differences across a deliberately coarse corporate versus independent divide.

Implementation detail. For all analyses, a creator’s segment is encoded as a binary indicator \(A_c\in\{0,1\}\) with \(A_c=1\) for agency-affiliated and \(A_c=0\) otherwise; for readability we report results using the labels “Agency” and “Independent,” treating the latter as a pseudo non-corporate segment used for segmentation rather than a coherent organizational portfolio.

All data analyzed in this study were drawn from publicly available online sources and contain no personally identifiable information beyond platform-assigned usernames. No interaction with human subjects was involved. According to the guidelines of the authors’ affiliated institutions, the project was exempt from formal ethics review.

Our empirical strategy has two layers built from the same deduplicated YouTube live chat logs. At the \emph{micro} level, we construct a fan--month panel that assumes a single \emph{oshi} per viewer per month, defined as the channel receiving the largest share of that viewer's chat activity (ties broken deterministically by channel ID). At the \emph{meso} level, we project the same logs into a creator--creator audience-overlap network that \emph{allows} viewers to interact with multiple channels in the same month; edges capture this multi-oshi activity as shared audiences. Separating these layers lets us link individual trajectories to network structure without mixing definitions, and keeps micro-level retention/switching comparable to meso-level cohesion.

\subsection{Retention Analysis (RQ1)}

This subsection addresses RQ1 by estimating short horizon retention, primary creator switching, and inactivity from a fan centered panel, and by comparing distributions across affiliations. We operationalize commitment at the fan level. A fan $u$ is \emph{active} in month $m$ if the fan posts at least one chat message on any creator’s channel. For fans active at $m$, the $k$ month retention indicator is
\[
\mathrm{ret}_{u}(m,k)=\mathbb{1}\{u \text{ is active in every month } m+1,\ldots,m+k\}.
\]

By requiring activity in every subsequent month, $\mathrm{ret}_{u}(m,k)$ follows a \emph{rolling retention} definition commonly used in product analytics.\footnote{Rolling retention counts users who are active in \emph{every} month of a horizon; in contrast, a laxer “return-by-$m{+}k$” definition counts users as retained if they return at least once within the $k$-month window.} Compared with the return-by-$m{+}k$ variant, this stricter continuity criterion yields lower absolute rates but avoids overcounting sporadic one-off activity and keeps retention aligned with our state taxonomy below (\textit{Same}/\textit{Retain}/\textit{Cross}/\textit{Drop}). We adopt it to reduce downward bias in churn estimates.

Segment level rates are the means of $\mathrm{ret}_{u}(m,k)$ over fans whose \emph{origin creator} $o_{u,m}$ at $m$ belongs to segment $s$. We report $k\in\{2,3\}$ by month, together with the monthly mean and an interquartile band from the 25th to the 75th percentiles (see the panels titled ``Two Month Retention'' and ``Three Month Retention'').

To characterize mobility within continued activity, we define each fan’s monthly \emph{oshi} (primary creator)
\[
o_{u,m}=\arg\max_{c} L_{u,c,m},
\]
where $L_{u,c,m}$ is the number of chat messages by $u$ on creator $c$ in month $m$ (ties are broken deterministically by lexicographic channel ID). The \emph{oshi switch} indicator equals $\mathbb{1}\{o_{u,m+1}\neq o_{u,m}\}$ among fans active in both $m$ and $m+1$. We summarize switch probabilities by affiliation and visualize their distributions across creators using violin plots. As a persistence proxy, we also compute the \emph{current run length} of each fan in month $m$, defined as the number of consecutive months of activity up to and including $m$, and we track its segment level mean over time (see ``Longest Run'').

To compare these commitment metrics between the Agency and Independent segments, we perform two sample tests on the distributions of creator level average scores. For each metric, we first assess normality for Agency and Independent using the Shapiro--Wilk test. If both groups are normal, we apply Welch $t$ tests with unequal variances allowed; otherwise, we use the nonparametric Mann--Whitney $U$ test. This two step procedure yields a robust comparison of central tendencies between the two creator populations.

\subsection{State-Transition Paths (RQ1)}

We further address RQ1 by decomposing multi step audience flows (Same, Retain, Cross, Drop) over a three month horizon, which complements the static metrics above. We define a fan’s origin in month $m$ as the pair $(c_0, s_0)$ of creator and affiliation, where $s_0 \in \{\textit{Agency}, \textit{Independent}\}$. For each of the subsequent three months $m+t$ with $t \in \{1,2,3\}$, we classify the fan into one of four mutually exclusive states based on chat activity:
\begin{itemize}
    \item \textit{Same}: The fan remains with the origin creator and posts at least one chat on channel $c_0$ in month $m+t$ (creator retention).
    \item \textit{Retain}: The fan does not engage with $c_0$ but engages with at least one other creator affiliated with the origin segment $s_0$ (segment retention).
    \item \textit{Cross}: The fan does not engage with any creator from $s_0$ but engages with at least one creator from the other segment (segment leakage).
    \item \textit{Drop}: The fan posts no chats on any channel we track in month $m+t$ (attrition).
\end{itemize}

For origins with $s_0=\textit{Independent}$, \textit{Retain} should be read as \emph{same-segment retention}: the fan remains active among independent creators, but not within a single organizational portfolio. Because “Independent” aggregates heterogeneous production models (from hobbyists to former corporate talents), this convention can slightly inflate segment-retention relative to a portfolio-based notion. We adopt it to keep the taxonomy symmetric across segments and to cleanly separate creator loyalty (\textit{Same}) from segment-level continuity (\textit{Retain}). Readers should interpret \textit{Independent}--\textit{Retain} conservatively; our claims emphasize relative patterns (e.g., \textit{Retain} vs.\ \textit{Drop}) rather than the absolute magnitude of this cell.\footnote{Figure labels are kept unchanged for comparability across months and segments. We further discuss this limitation in the text.} Formally, segment membership in these transitions is determined by the binary indicator \(A_c\); “Independent” corresponds to the complement \(A_c=0\).

When a month contains multiple categories of activity, we apply a strict hierarchy: \textit{Same} $>$ \textit{Retain} $>$ \textit{Cross} $>$ \textit{Drop}. We estimate the share of fans in each state for $t\in\{1,2,3\}$ conditional on $s_0$ and visualize the resulting trajectories using Sankey diagrams (see Fig.~2). This construction separates creator loyalty (\textit{Same}) from portfolio stabilization (\textit{Retain}) and measures churn explicitly (\textit{Cross}, \textit{Drop}), which allows us to test whether affiliation fosters movement within a portfolio and dampens exit.

\subsection{Macro-Network Evolution (RQ2)}

This subsection addresses RQ2 by constructing monthly audience overlap networks and by comparing meso level structure between affiliated and independent subgraphs. For each month $m$, we build a creator to creator network by projecting a bipartite fan to creator graph. Let $U_{c,m}$ be the set of unique users who posted at least one chat message on creator $c$ in month $m$. For any pair of active creators $(i,j)$, we quantify audience overlap using the Simpson coefficient
\[
S_{ij,m}=\frac{|U_{i,m}\cap U_{j,m}|}{\min\{|U_{i,m}|,|U_{j,m}|\}},
\]
which is robust to asymmetries in audience size. We retain an undirected edge $(i,j)$ in month $m$ if two criteria are met: (1) $S_{ij,m}\ge\theta$ and (2) the number of shared viewers $|U_{i,m}\cap U_{j,m}|$ is at least $n_{\min}$.\footnote{The primary analysis uses $\theta=0.05$ and $n_{\min}=10$. Findings are qualitatively stable for $\theta \in [0.03,0.07]$ and $n_{\min} \in \{5,15\}$.} The monthly graph is $G_m=(V_m,E_m)$, where the node set $V_m$ includes only creators with $|U_{c,m}|\ge\tau_U$ unique chatters, which excludes inactive channels.\footnote{We set $\tau_U=25$; results are consistent for $\tau_U\in\{10,50\}$.}

To compare organizational regimes, we analyze subgraphs $G_m^{(s)}$ for $s\in\{\textit{Agency},\textit{Independent}\}$ by restricting $V_m$ and $E_m$ to nodes within each segment, which isolates within segment cohesion from bridges between segments.

Because the number of active nodes $|V_m^{(s)}|$ varies by month, we employ two metrics that account for scale. First, the average degree
\[
\bar{k}_{s,m}=\frac{1}{|V_m^{(s)}|}\sum_{v\in V_m^{(s)}} \deg_{G_m^{(s)}}(v)
\]
measures local connectedness and is comparatively less sensitive to network size. Second, the edge density
\[
\delta_{s,m}=\frac{2|E_m^{(s)}|}{|V_m^{(s)}|(|V_m^{(s)}|-1)}
\]
quantifies global saturation. Reporting both helps distinguish structural weakening from dilution driven by size and allows detection of localized clustering even when global densities converge.

For formal inference over time, we use paired tests on monthly differences between Agency and Independent. We first assess normality of the paired differences with the Shapiro--Wilk test. If normality holds with $p\ge 0.05$, we apply a paired samples $t$ test; otherwise, we use the Wilcoxon signed rank test.

\subsection{Representative Topologies (RQ2)}

We complement RQ2 with representative snapshots that make meso level differences interpretable. For selected months, for example January 2023, January 2024, and January 2025, we visualize $G_m^{(\textit{Agency})}$ and $G_m^{(\textit{Independent})}$ constructed with the same $\theta$, $n_{\min}$, and $\tau_U$. Layouts use a force directed algorithm with a fixed random seed and constant hyperparameters to preserve comparability across segments within a month and to reduce visual drift across years.

Visual encoding follows fixed conventions. Node size scales with degree within the subgraph to highlight hubs. Edge width is log scaled by $S_{ij,m}$, so stronger ties appear thicker; edges use uniform partial transparency to mitigate clutter. Isolated nodes within each subgraph are omitted for readability. These choices are intended to ensure that visible patterns reflect genuine organizational differences rather than artifacts of styling.

In addition to these segment-specific monthly graphs, we also construct a unified audience–overlap network that pools all months into a single weighted graph $G^{(\textit{unified})} = (V, E^{(\textit{unified})})$. The vertex set $V$ contains all creators appearing in any month. For each unordered pair $(i,j)$, we aggregate monthly Simpson weights as
\[
w_{ij} = \sum_m S_{ij,m},
\]
and include an edge in $E^{(\textit{unified})}$ whenever $w_{ij} > 0$. On this unified graph we compute node-level measures that summarize how each creator is positioned in the overall audience ecosystem. Specifically, we use (i) weighted degree $d_i^{(w)} = \sum_j w_{ij}$, (ii) unweighted degree $d_i$ (the number of neighbors), (iii) the standard local clustering coefficient $C_i$ based on triangle density in the unweighted projection, and (iv) unweighted betweenness centrality defined over shortest paths.

To examine how organizational affiliation shapes these positions, we stratify these node-level metrics by the binary affiliation indicator $A_c \in \{\textit{Agency}, \textit{Independent}\}$. For each metric we compare the creator-level distributions between agency-affiliated and independent VTubers using violin plots with overlaid points and a horizontal line indicating the mean. This unified-graph analysis complements the segment-specific snapshots by showing that affiliation is associated not only with within-portfolio cohesion but also with systematically different positions in the audience-overlap network as a whole.

\section{Results}\label{sec:res}

\subsection{Retention Analysis (RQ1)}

\begin{figure}[!t]
  \centering

  \subfloat[Two month retention\label{fig:2m_comp}]{
    \includegraphics[width=0.47\linewidth]{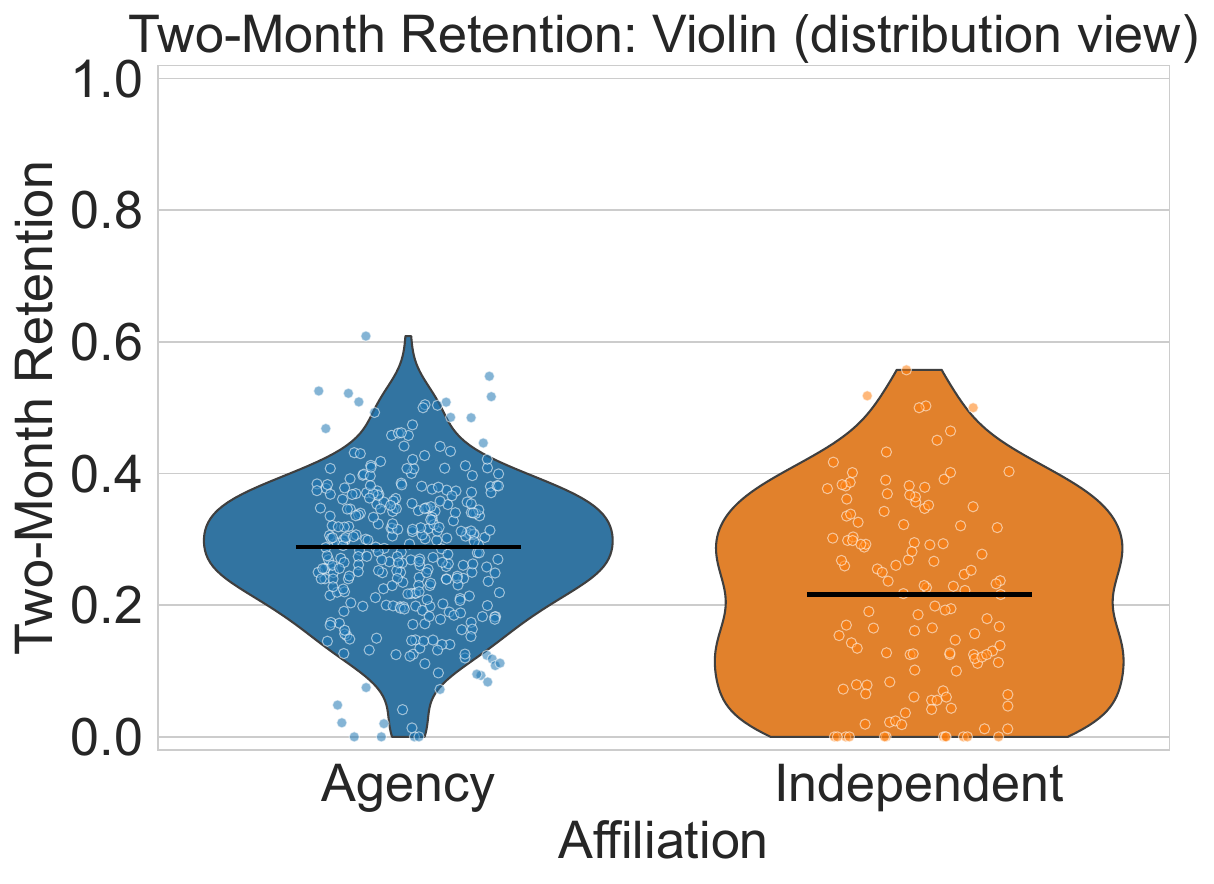}
  }%
  \hfill
  \subfloat[Three month retention\label{fig:3m_comp}]{
    \includegraphics[width=0.47\linewidth]{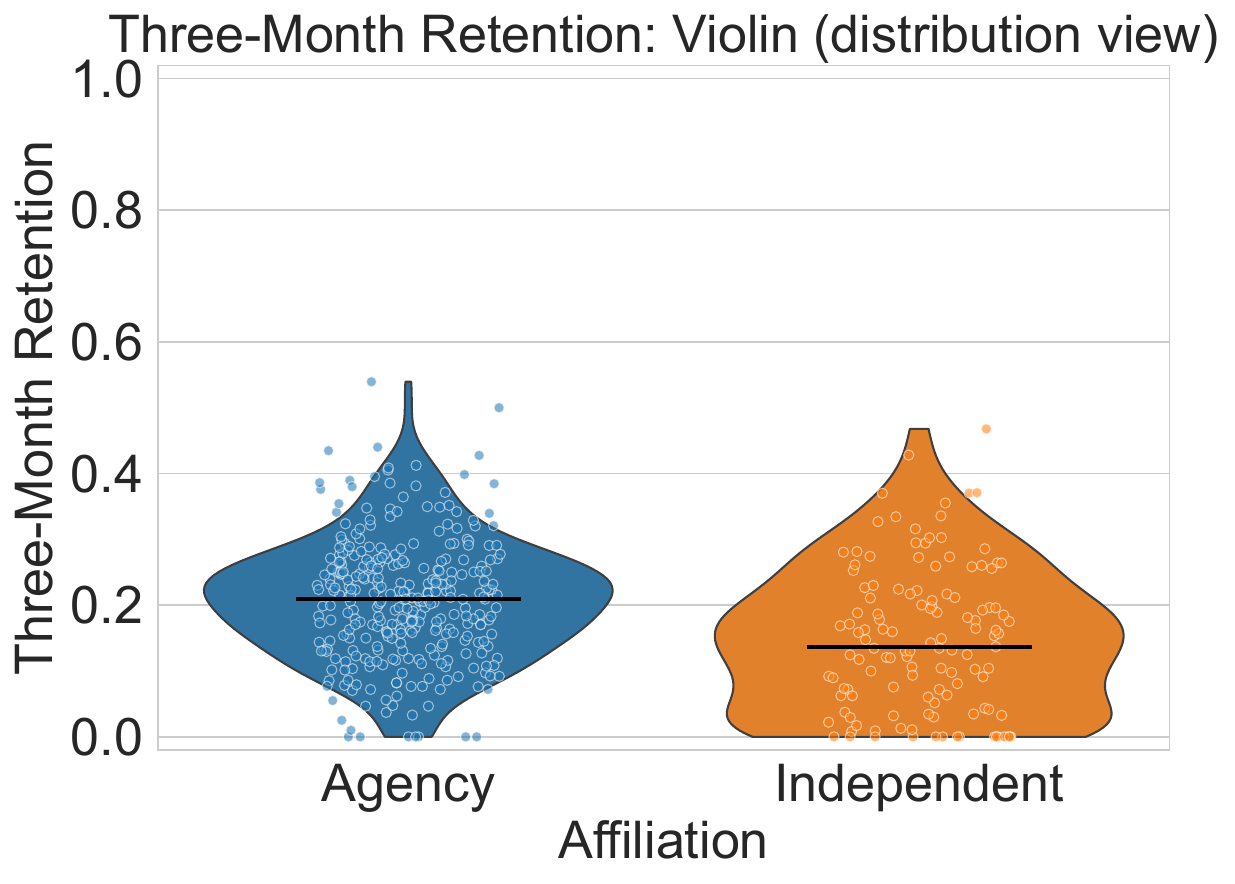}
  }\\[-0.6\baselineskip]%

  \subfloat[Longest active run\label{fig:longest_run_comp}]{
    \includegraphics[width=0.47\linewidth]{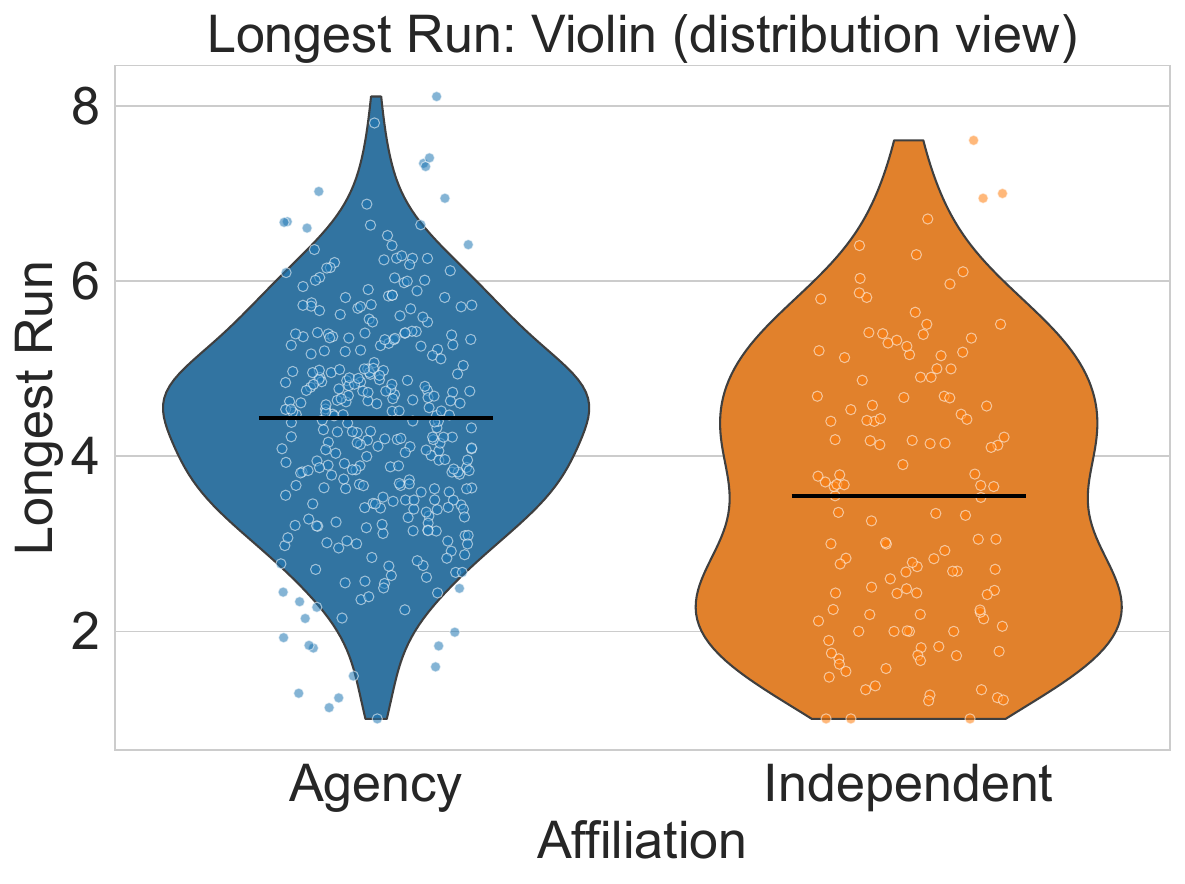}
  }%
  \hfill
  \subfloat[Oshi switch probability\label{fig:oshi_change_comp}]{
    \includegraphics[width=0.47\linewidth]{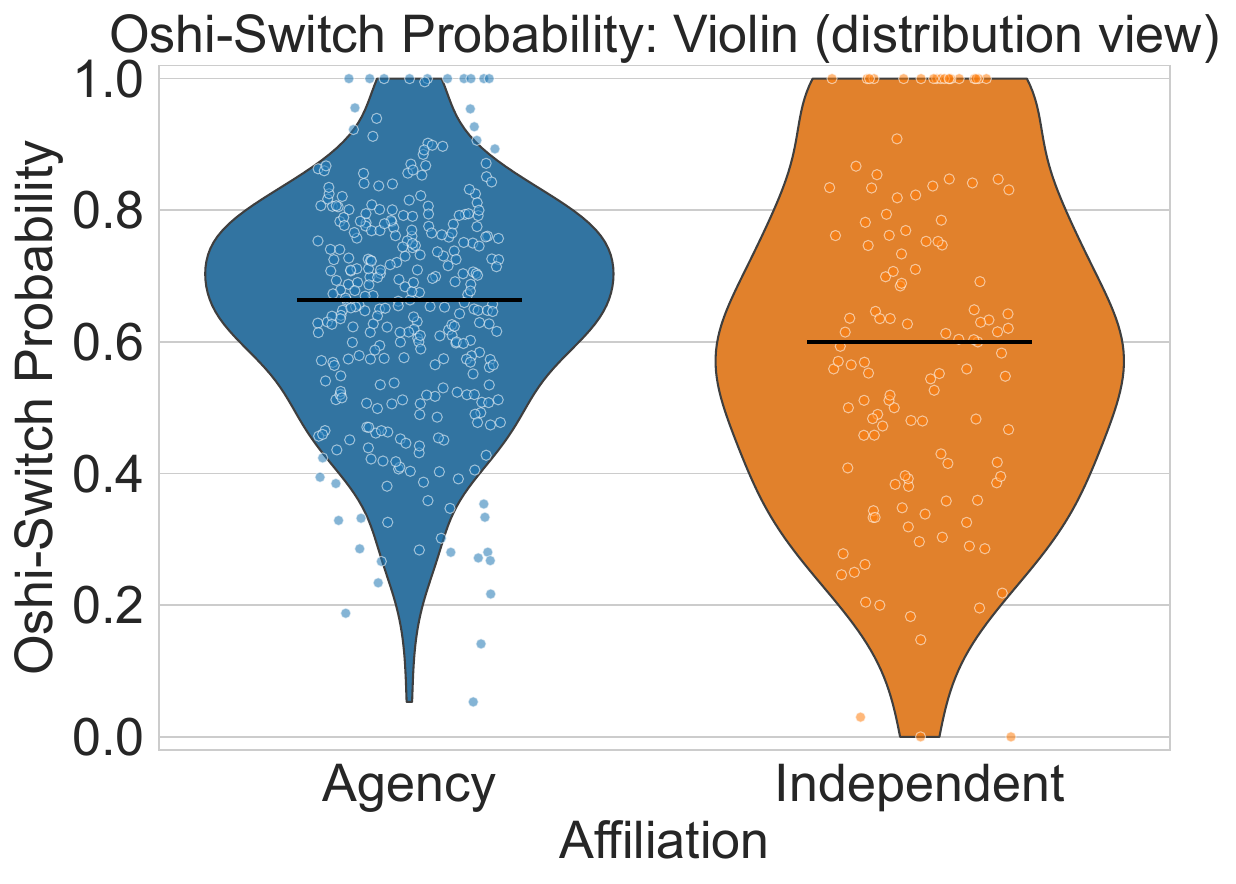}
  }

  \caption{Commitment metrics by affiliation (Agency vs.\ Independent). Each panel shows a violin plot. Metrics: (a) two month retention; (b) three month retention; (c) longest active run; (d) oshi switch probability.}
  \label{fig:commitment_comparison_grid}
\end{figure}

Analysis of fan commitment metrics reveals clear differences by creator affiliation. Fans originating from agency affiliated creators show significantly higher two and three month retention rates than those from the independent cohort (\autoref{fig:2m_comp}, \autoref{fig:3m_comp}). The same pattern appears in the longest active run, which is higher for the agency group (\autoref{fig:longest_run_comp}). As detailed in \autoref{tab:commitment_tests}, these differences are statistically significant based on the Mann--Whitney $U$ test ($p<0.001$).

Higher persistence does not imply loyalty to a single creator. The probability of an \emph{oshi} (primary creator) switch is also significantly higher within the agency segment (\autoref{fig:oshi_change_comp}; \autoref{tab:commitment_tests}). This combination, namely greater overall retention together with more frequent creator switching, characterizes the agency cohort.

These differences in commitment and mobility were not transient. As shown in \autoref{tab:commitment_tests}, gaps between the agency and independent segments persisted throughout the observation period, which indicates systematic differences in engagement trajectories by affiliation. Where reported, rank biserial correlations ($r \approx 0.22$ to $0.29$) suggest small to moderate effects.

\begin{table}[t]
\centering
\caption{Statistical comparison of commitment metrics (Agency vs.\ Independent).}
\label{tab:commitment_tests}
\scriptsize
\begingroup
\setlength{\tabcolsep}{4pt}
\renewcommand{\arraystretch}{1.1}

\begin{tabular}{l r r r}
\multicolumn{4}{c}{\textit{(a) Descriptive means}}\\
\hline
Metric & Mean A & Mean I & $\Delta$ \\
\hline
Two Month Retention     & 0.613 & 0.509 & 0.104 \\
Three Month Retention   & 0.500 & 0.388 & 0.112 \\
Longest Active Run      & 5.106 & 4.197 & 0.909 \\
Oshi Switch Probability & 0.597 & 0.490 & 0.107 \\
\hline
\end{tabular}

\medskip

\textit{(b) Inferential tests for group differences}
\resizebox{\columnwidth}{!}{
\begin{tabular}{l l l r c}
\hline
Metric & Test & Statistic & $p$ & $r$ \\
\hline
Two Month Retention       & Mann--Whitney $U$ & $U{=}20117.0$ & $<0.001$ & 0.288 \\
Three Month Retention     & Mann--Whitney $U$ & $U{=}19895.0$ & $<0.001$ & 0.274 \\
Longest Active Run        & Mann--Whitney $U$ & $U{=}19163.0$ & $0.001$  & 0.222 \\
Oshi Switch Probability   & Mann--Whitney $U$ & $U{=}19283.5$ & $0.001$  & 0.230 \\
\hline
\end{tabular}
}
\par\medskip
\begin{minipage}{\columnwidth}\footnotesize
Notes: Means are per creator averages; $\Delta$ is defined as Agency minus Independent. $n_A$ and $n_I$ denote the numbers of creators in the Agency and Independent groups. Test choice was based on normality assessed with the Shapiro--Wilk test. Because normality was not met for all groups, the nonparametric Mann--Whitney $U$ test was used. Effect size is reported as the rank biserial correlation $r$.
\end{minipage}
\endgroup
\end{table}

\subsection{State-Transition Paths (RQ1)}

\begin{figure}[!t]
  \centering
  \subfloat[Agency origin\label{fig:sankey_agency}]{
    \includegraphics[width=0.95\linewidth]{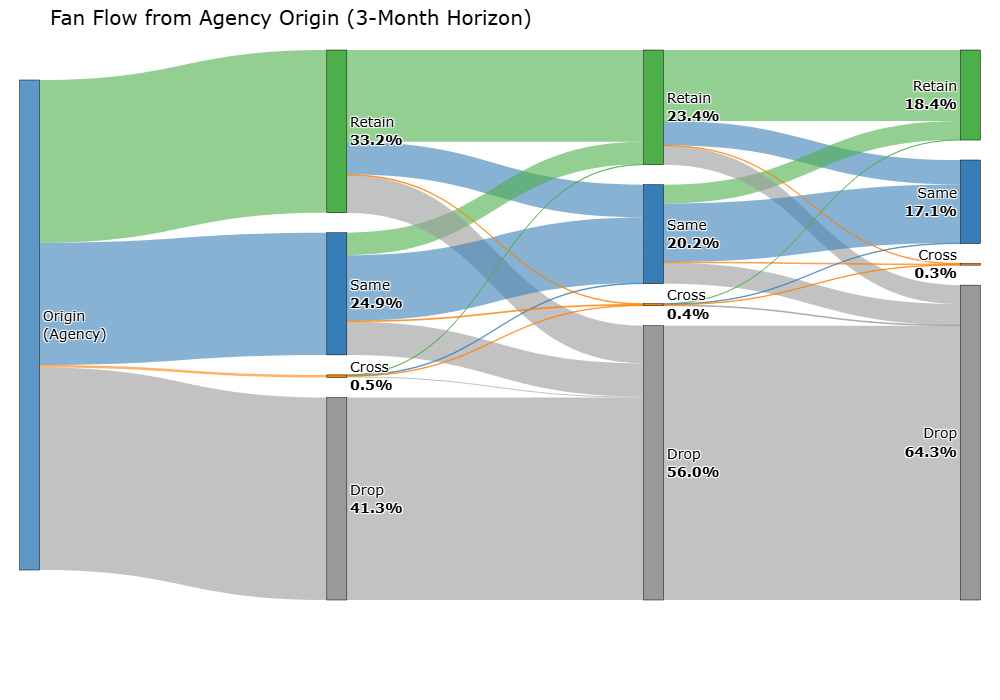}
  }\par\vspace{-0.4\baselineskip}
  \subfloat[Independent origin\label{fig:sankey_independent}]{
    \includegraphics[width=0.95\linewidth]{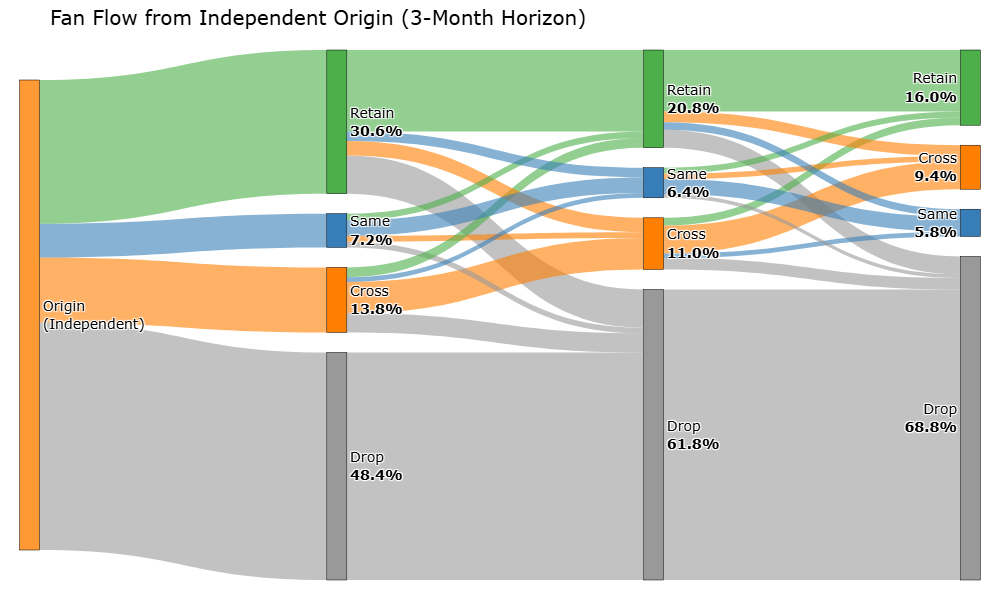}
  }
  \caption{Sankey diagrams of fan flows over a three month horizon for cohorts from (a) agency and (b) independent origins. At each monthly time point (T1 to T3), fans are classified into one of four states: \textit{Same}, \textit{Retain}, \textit{Cross}, or \textit{Drop}.}
  \label{fig:sankey_flow_comparison}
\end{figure}

To clarify pathways of fan mobility, we decomposed multi step audience flows and visualized where active fans migrate over a three month horizon (\autoref{fig:sankey_flow_comparison}). Initial one month transitions show a pronounced effect of retention within the segment for agency origin cohorts: fans are more likely to remain with the same creator (\textit{Same}) or to engage another creator within the same segment (\textit{Retain}), while migration to the independent segment (\textit{Cross}) is limited by comparison. In contrast, independent origin cohorts show a larger share of attrition in the next month (\textit{Drop}) and only modest movement into the agency segment.

Beyond the first month, the three step flows reveal a recurrent pattern of circulation that is more pronounced in the agency segment. As shown in \autoref{fig:sankey_flow_comparison}, agency origin cohorts display a visible ``circulate and recapture'' dynamic: fans who leave a focal creator often reappear at other affiliated channels in a subsequent month, and some later return to their origin. This internal recirculation is much less evident among independent origin cohorts, who are more likely to become inactive after disengaging from their initial creator.

Taken together, these flow dynamics are consistent with a mechanism in which portfolios retain fans within the segment rather than losing them to external churn, which provides a plausible explanation for the commitment patterns above.

\subsection{Macro-Network Evolution (RQ2)}

\begin{figure}[!t]
  \centering
  \subfloat[Network density\label{fig:density}]{
    \includegraphics[width=0.95\linewidth]{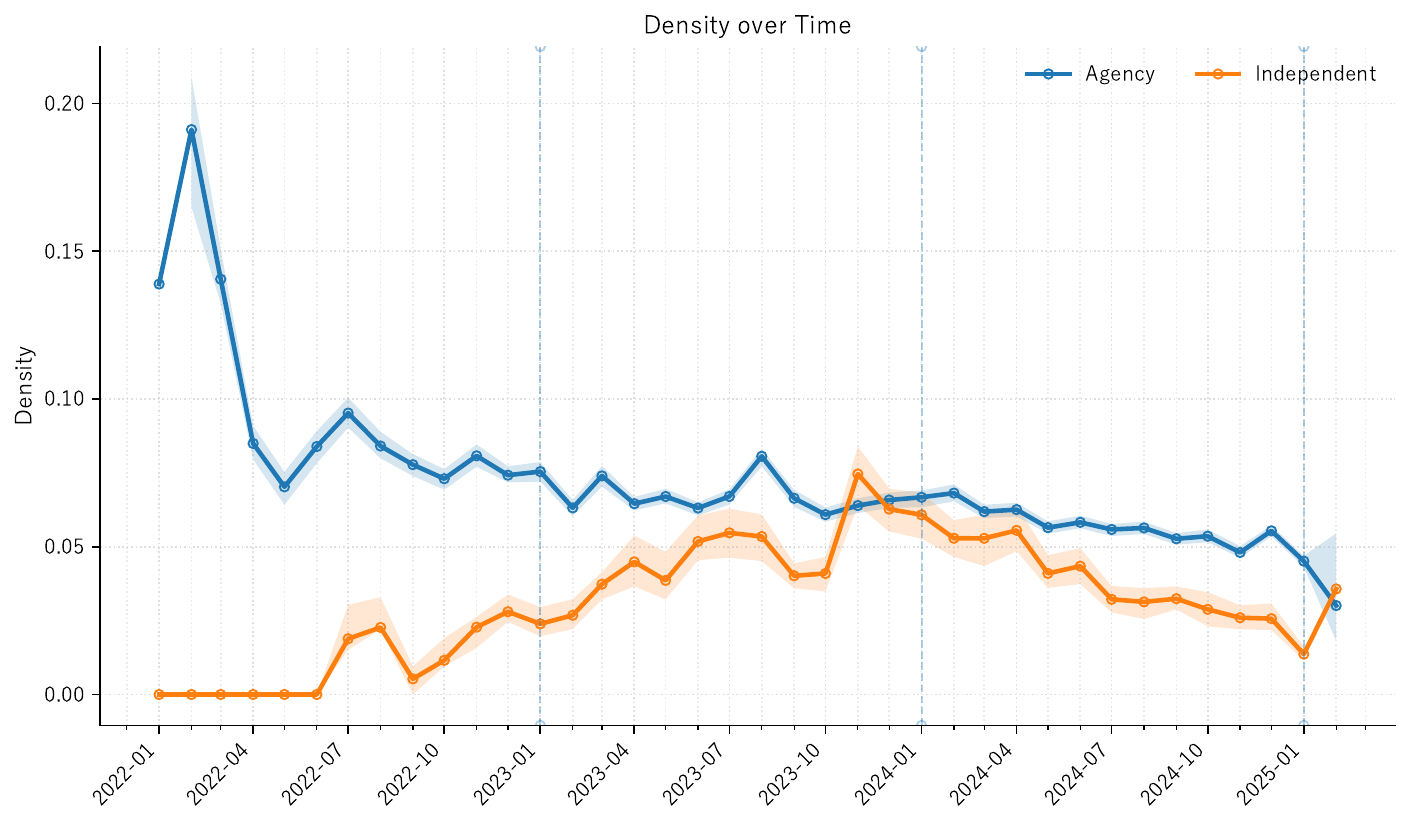}
  }\par
  \subfloat[Average degree\label{fig:avg_degree}]{
    \includegraphics[width=0.95\linewidth]{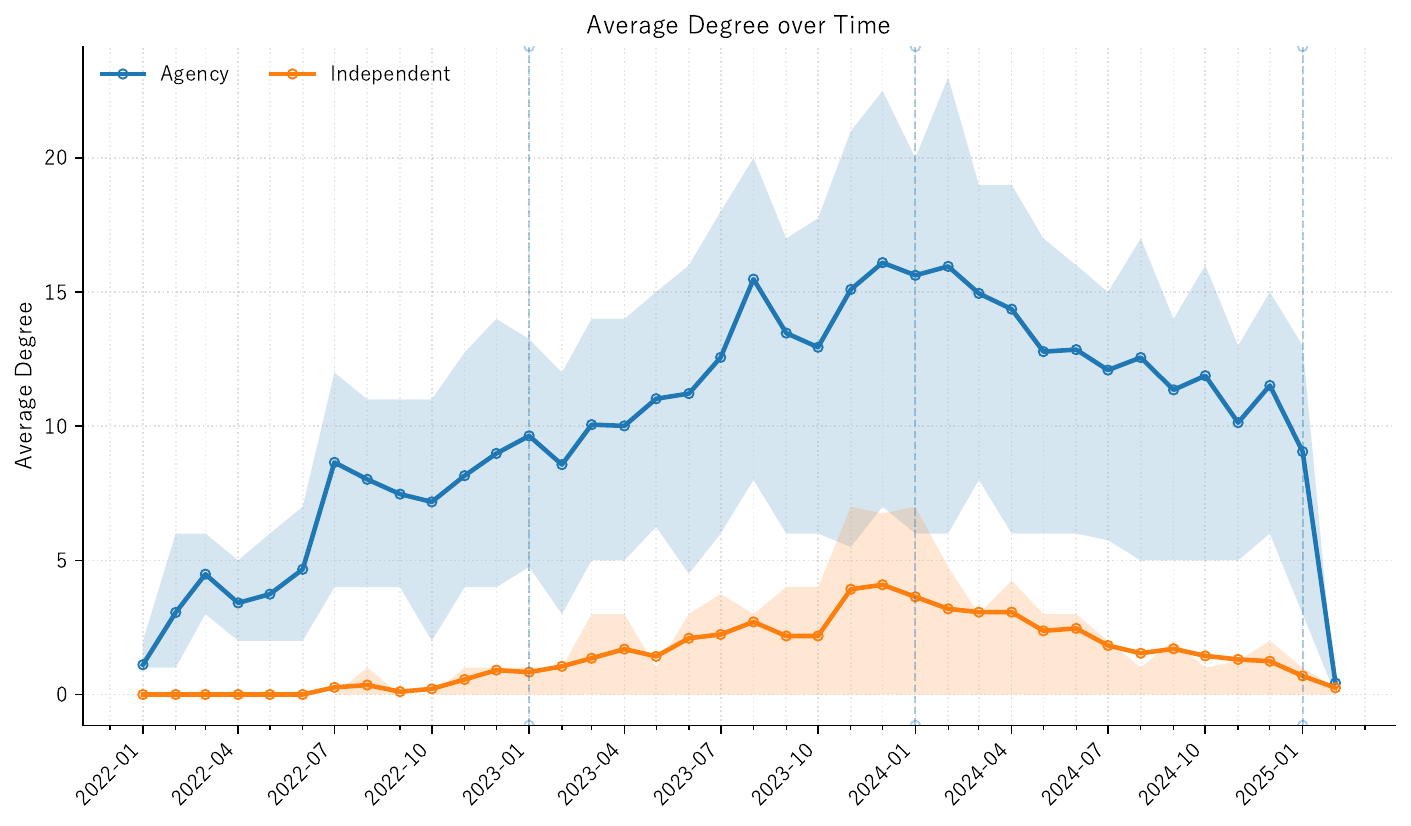}
  }
  \caption{Time series of network metrics for the Agency and Independent segments. (a) Network density and (b) average degree over time.}
  \label{fig:network_metrics_timeseries_vertical}
\end{figure}

The temporal evolution of the creator network reveals distinct structural patterns between the Agency and Independent segments. A key dynamic is the decoupling of global saturation and local connectivity (\autoref{fig:network_metrics_timeseries_vertical}). While the overall edge density of both segments converges toward similar levels by late 2024, which indicates a general broadening of audience overlap across the ecosystem (\autoref{fig:density}), the average degree follows a different trend. The Agency segment establishes and progressively widens its advantage in average degree, which is consistent with the formation of denser and more tightly knit local neighborhoods around affiliated creators (\autoref{fig:avg_degree}). This divergence aligns with selective connectivity within portfolios rather than indiscriminate tie formation.

Monthly paired comparisons support these structural disparities. As detailed in \autoref{tab:macro_network_tests}, the Agency segment exhibits significantly higher values across multiple metrics, including the number of nodes and edges, network density, and the size of the largest connected component (Wilcoxon signed rank, $p<0.001$ for all). Metrics of local structure, namely average degree and the clustering coefficient, show particularly large and significant differences (paired samples $t$ test for clustering, $t=20.04$, $p<0.001$, $d=3.25$). Taken together, these findings are unlikely to arise from random fluctuation and are consistent with coordinated practices at the portfolio level, such as cross promotions and synchronized events, that selectively thicken ties among a curated subset of creators.

\begin{table}[t]
\centering
\caption{Paired monthly network metrics (Agency vs.\ Independent).}
\label{tab:macro_network_tests}
\scriptsize
\begingroup
\setlength{\tabcolsep}{3pt}
\renewcommand{\arraystretch}{1.1}

\textit{(a) Descriptive means}\\[-0.25ex]
\begin{tabular}{l r r r r}
\hline
Metric & $n$ & Mean A & Mean I & $\Delta$ \\
\hline
Nodes $|V|$               & 38 & 155.74 & 38.58 & 117.16 \\
Edges $|E|$               & 38 & 926.71 & 38.21 & 888.50 \\
Density $\delta$          & 38 & 0.073  & 0.031 & 0.042  \\
Avg.\ degree $\bar{k}$    & 38 & 10.018 & 1.473 & 8.545  \\
Clustering coefficient    & 38 & 0.639  & 0.185 & 0.454  \\
Largest connected component & 38 & 116.32 & 10.53 & 105.79 \\
\hline
\end{tabular}

\medskip

\textit{(b) Inferential tests for paired differences}\\[-0.25ex]
\resizebox{\columnwidth}{!}{
\begin{tabular}{l r r l l r c}
\hline
Metric & $n$ & $p_{\mathrm{SW}}$ & Test & Statistic & $p$ & $d$ \\
\hline
Nodes $|V|$               & 38 & 0.001     & Wilcoxon          & $W{=}0$      & $<0.001$ & NA  \\
Edges $|E|$               & 38 & 0.030     & Wilcoxon          & $W{=}0$      & $<0.001$ & NA  \\
Density $\delta$          & 38 & $<0.001$  & Wilcoxon          & $W{=}8$      & $<0.001$ & NA  \\
Avg.\ degree $\bar{k}$    & 38 & 0.004     & Wilcoxon          & $W{=}0$      & $<0.001$ & NA  \\
Clustering coefficient    & 38 & 0.060     & paired $t$ test   & $t{=}20.04$  & $<0.001$ & 3.25 \\
Largest connected component & 38 & 0.028   & Wilcoxon          & $W{=}0$      & $<0.001$ & NA  \\
\hline
\end{tabular}
}
\par\medskip
\begin{minipage}{\columnwidth}\footnotesize
Notes: Means are monthly; $\Delta$ is defined as Agency minus Independent. $n$ is the number of months with paired observations. $p_{\mathrm{SW}}$ is the $p$ value from the Shapiro--Wilk test for paired differences. If $p_{\mathrm{SW}}\!\ge\!0.05$, a paired $t$ test is used and Cohen’s $d$ is shown; otherwise the Wilcoxon signed rank test is used (statistic $W$). Values $p<0.001$ are shown as ``$<0.001$''.
\end{minipage}
\endgroup
\end{table}

\subsection{Representative Topologies (RQ2)}

\begin{figure}[!t]
  \centering

  \subfloat[2023 Independent\label{fig:net_2023_ind}]{
    \includegraphics[width=0.47\linewidth,trim=6pt 6pt 6pt 6pt,clip]{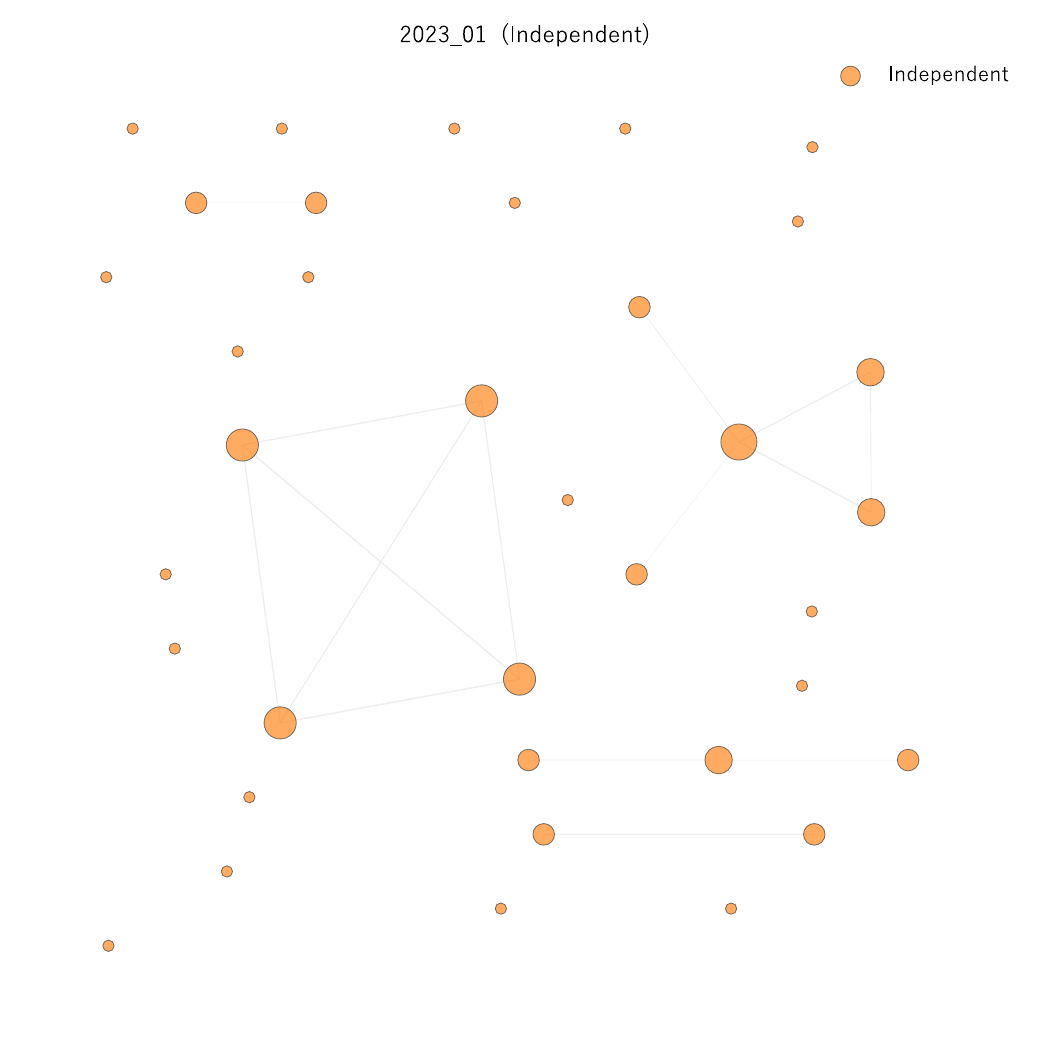}
  }%
  \hfill
  \subfloat[2023 Agency\label{fig:net_2023_ag}]{
    \includegraphics[width=0.47\linewidth,trim=6pt 6pt 6pt 6pt,clip]{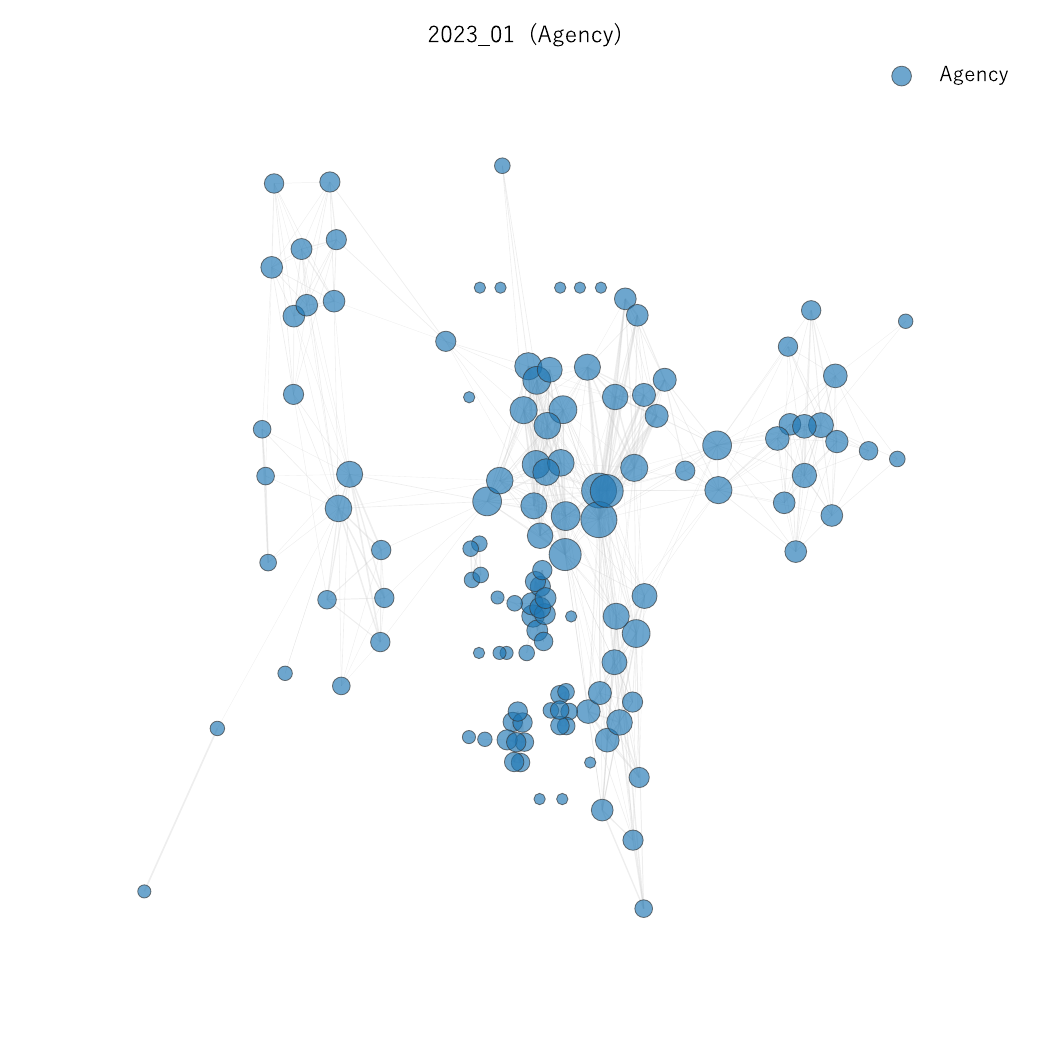}
  }\\[-0.6\baselineskip]%

  \subfloat[2024 Independent\label{fig:net_2024_ind}]{
    \includegraphics[width=0.47\linewidth,trim=6pt 6pt 6pt 6pt,clip]{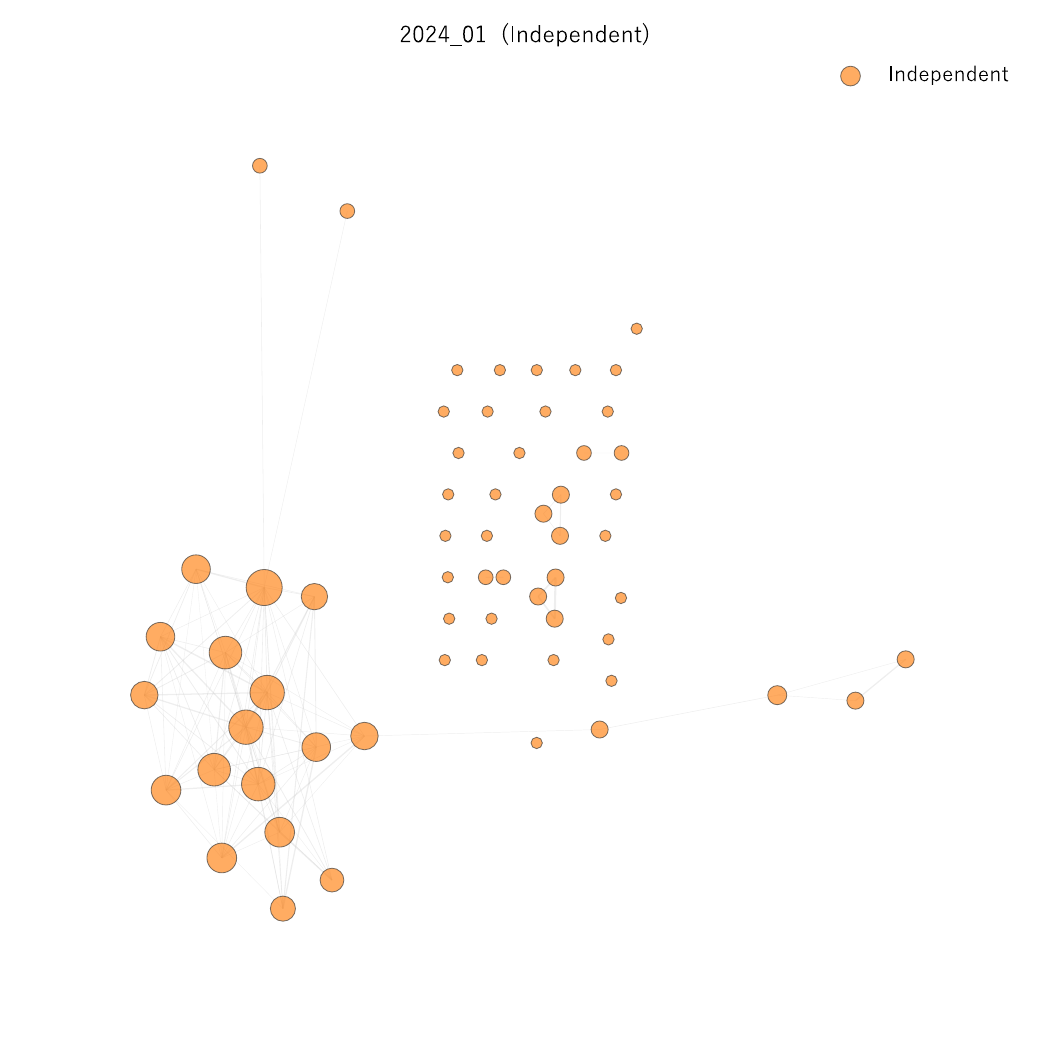}
  }%
  \hfill
  \subfloat[2024 Agency\label{fig:net_2024_ag}]{
    \includegraphics[width=0.47\linewidth,trim=6pt 6pt 6pt 6pt,clip]{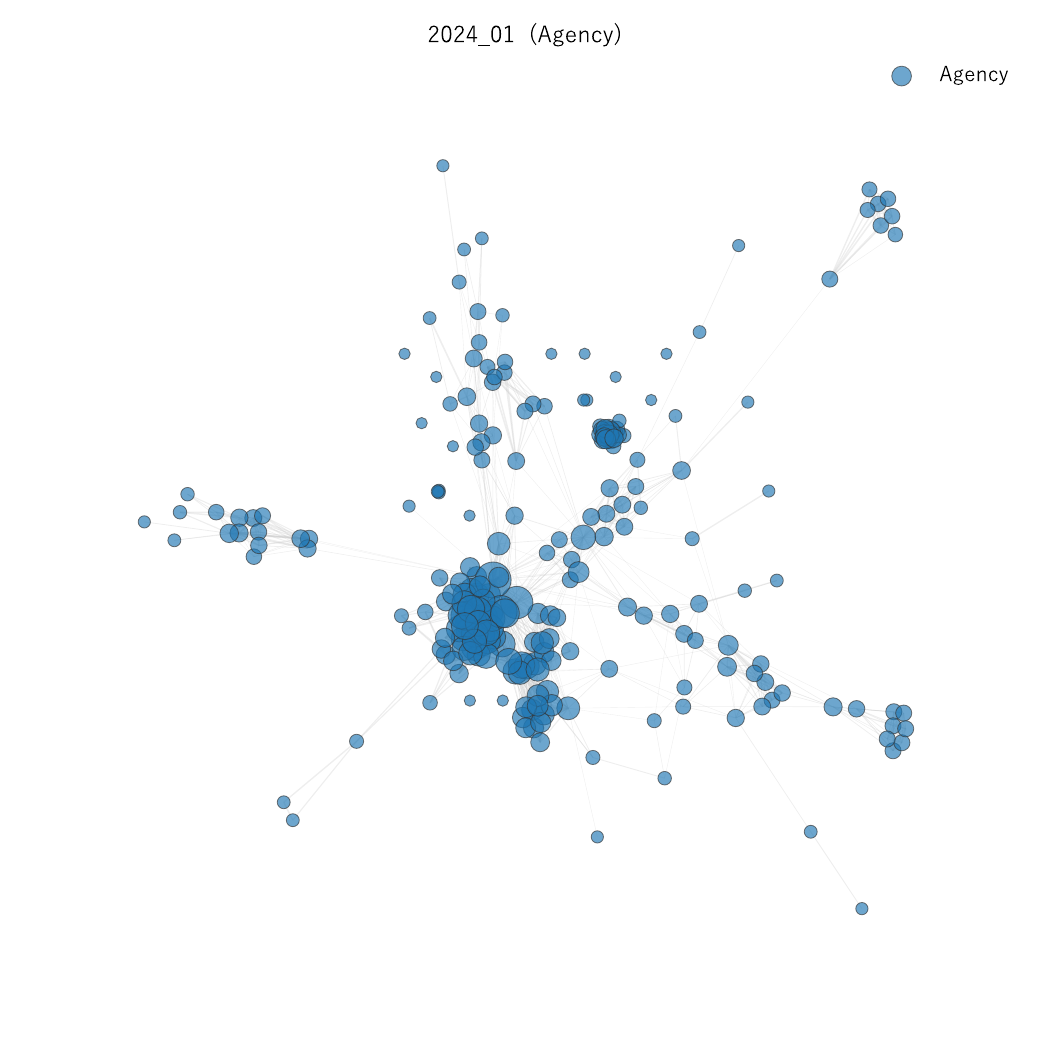}
  }\\[-0.6\baselineskip]%

  \subfloat[2025 Independent\label{fig:net_2025_ind}]{
    \includegraphics[width=0.47\linewidth,trim=6pt 6pt 6pt 6pt,clip]{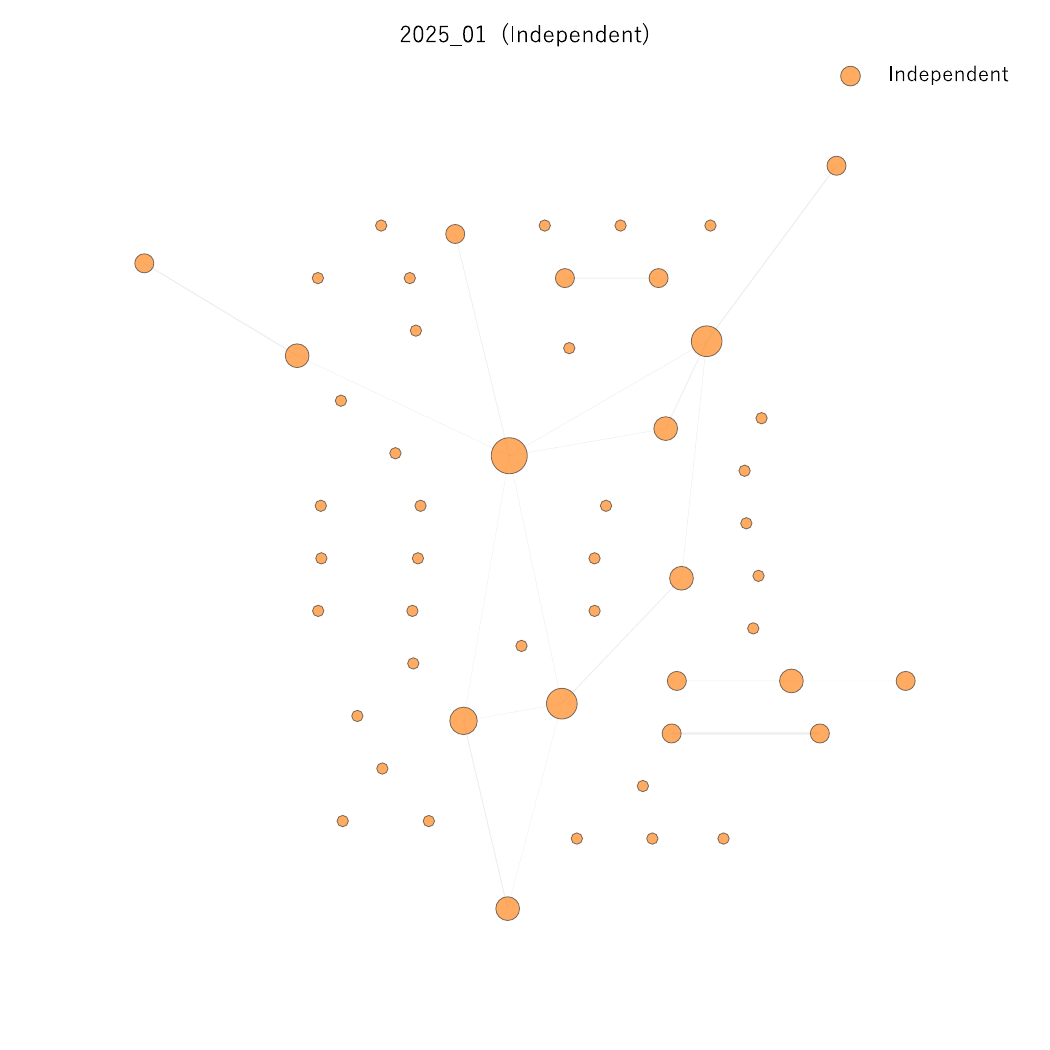}
  }%
  \hfill
  \subfloat[2025 Agency\label{fig:net_2025_ag}]{
    \includegraphics[width=0.47\linewidth,trim=6pt 6pt 6pt 6pt,clip]{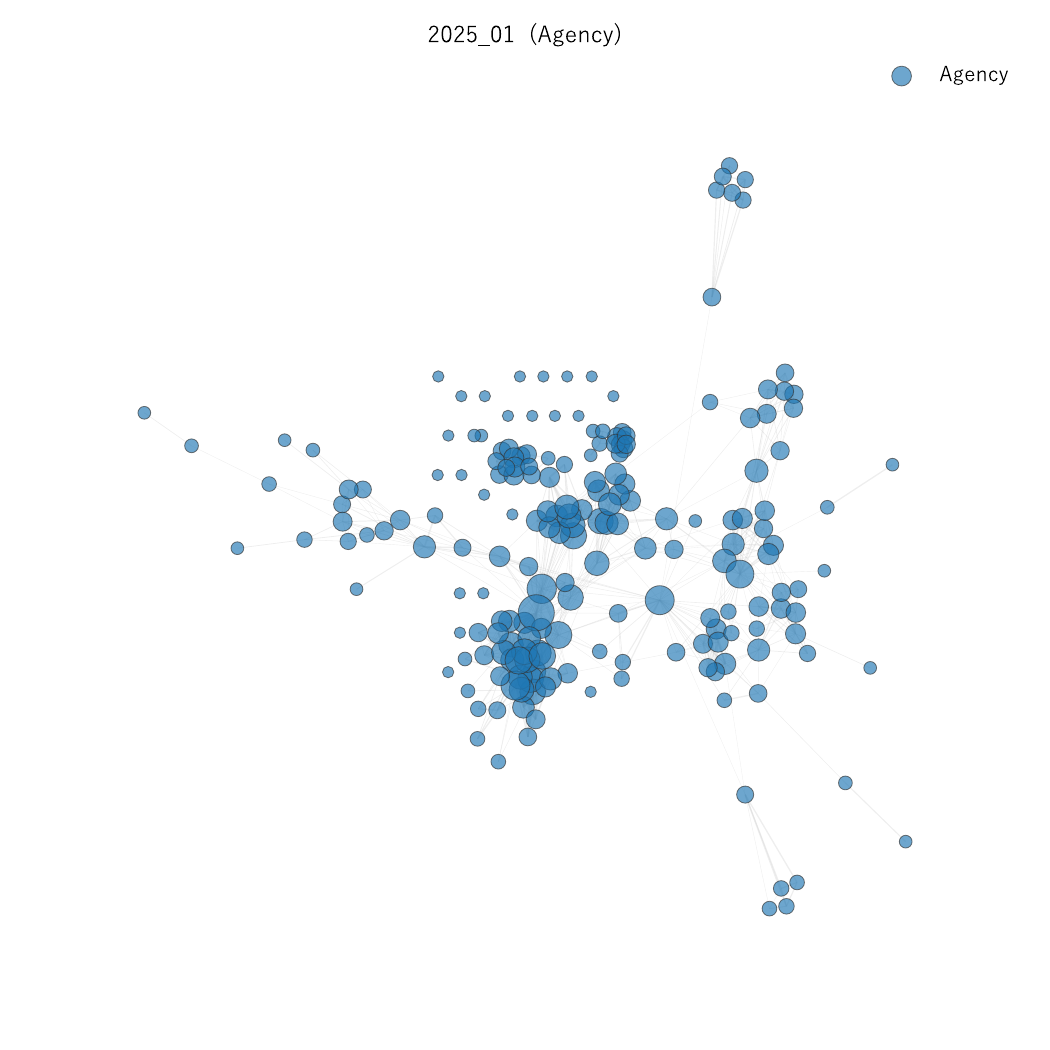}
  }

  \caption{Representative snapshots (January of each year) for Independent and Agency segments. Force directed layouts with fixed seeds; node size $\propto$ degree, edge width $\propto S_{ij,m}$. Isolated nodes are omitted for readability.}
  \label{fig:vtuber_networks}
\end{figure}

Representative snapshots make the meso level differences interpretable (\autoref{fig:vtuber_networks}). Agency subgraphs consistently concentrate into multiple well connected hubs linked by short bridges, forming dense cores with redundant paths. Independent subgraphs appear more fragmented, with more peripheral nodes, thinner bridges, and smaller local clusters. These visual patterns align with the quantitative metrics: Agency networks develop denser \emph{local} neighborhoods (higher average degree) without a corresponding rise in \emph{global} saturation (density), a configuration consistent with coordinated proximity within the segment.

Substantively, the hub and bridge geometry observed for the Agency segment is consistent with portfolio practices such as cross promotional appearances, synchronized schedules, and shared events that internalize audience spillover. Such structure provides fans with multiple and easily accessible alternatives within the portfolio, which facilitates circulation and potential recapture. In contrast, Independent snapshots lack cohesive cores and offer fewer redundant pathways, which implies fewer opportunities to recapture fans who drift from a specific creator.

\begin{figure}[!t]
  \centering
  \subfloat[Weighted degree\label{fig:unified_deg_weighted}]{
    \includegraphics[width=0.47\linewidth]{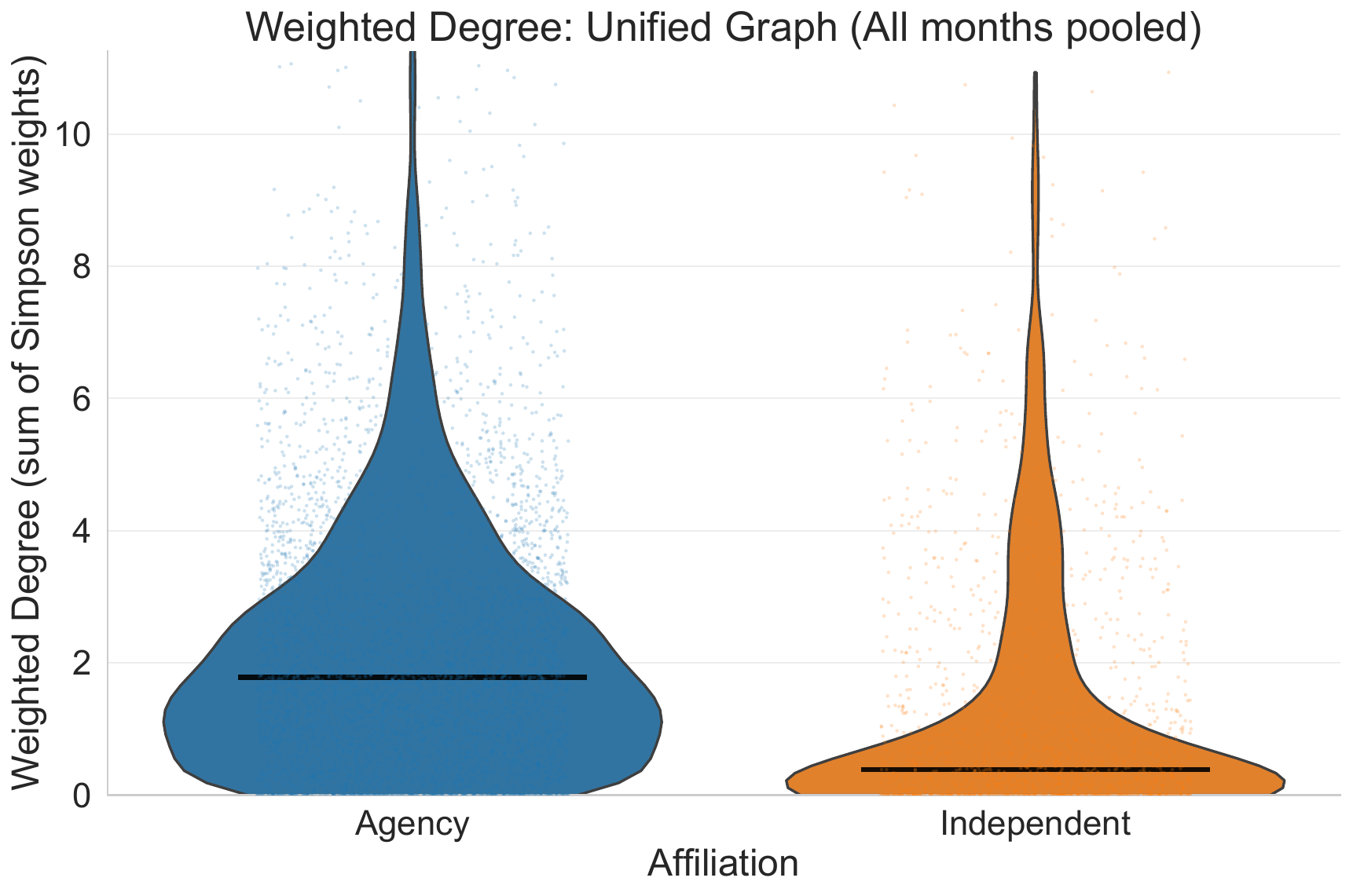}
  }%
  \hfill
  \subfloat[Degree\label{fig:unified_deg}]{
    \includegraphics[width=0.47\linewidth]{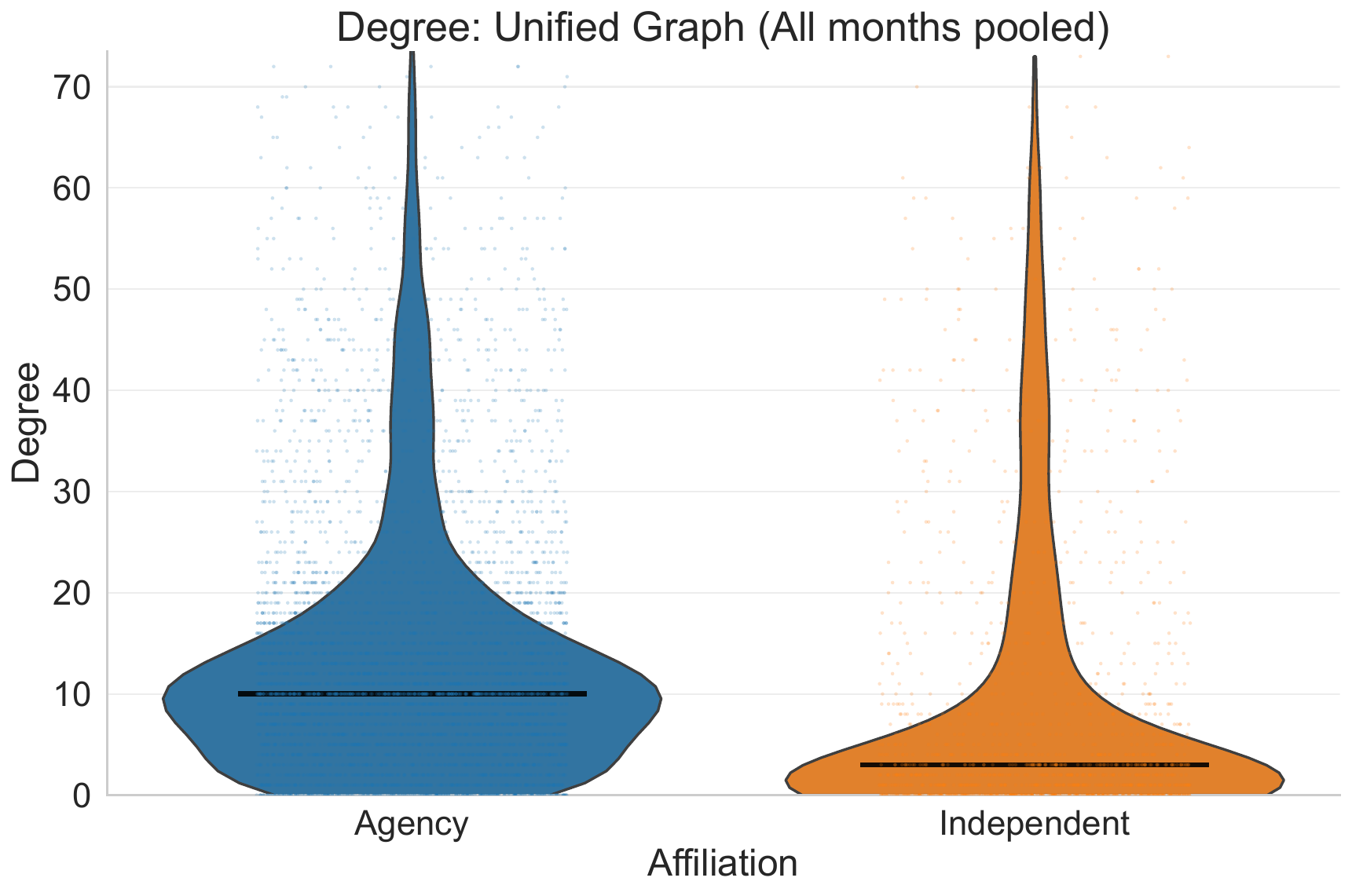}
  }\\[-0.6\baselineskip]%

  \subfloat[Clustering coefficient\label{fig:unified_clust}]{
    \includegraphics[width=0.47\linewidth]{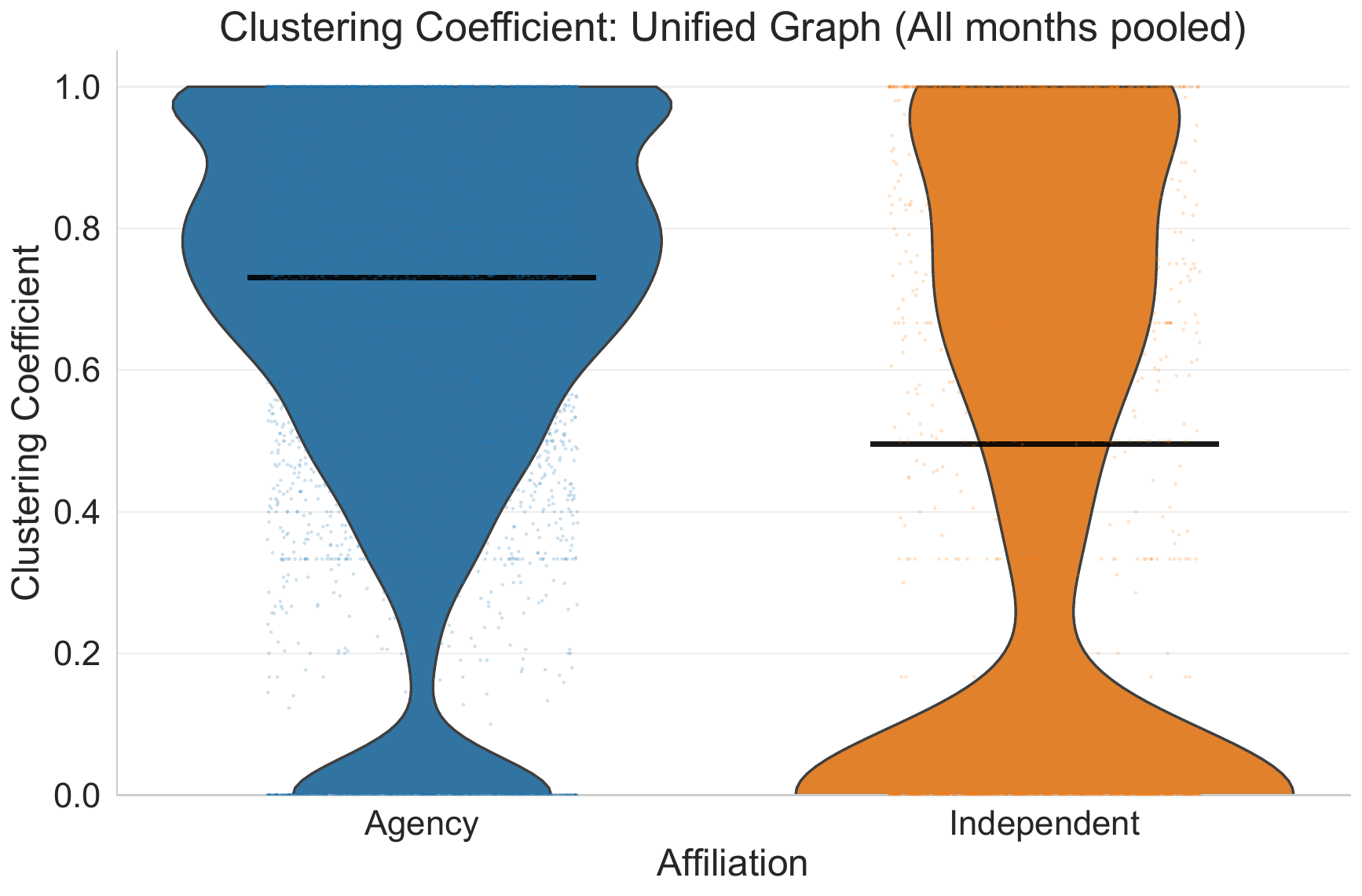}
  }%
  \hfill
  \subfloat[Betweenness centrality\label{fig:unified_betw}]{
    \includegraphics[width=0.47\linewidth]{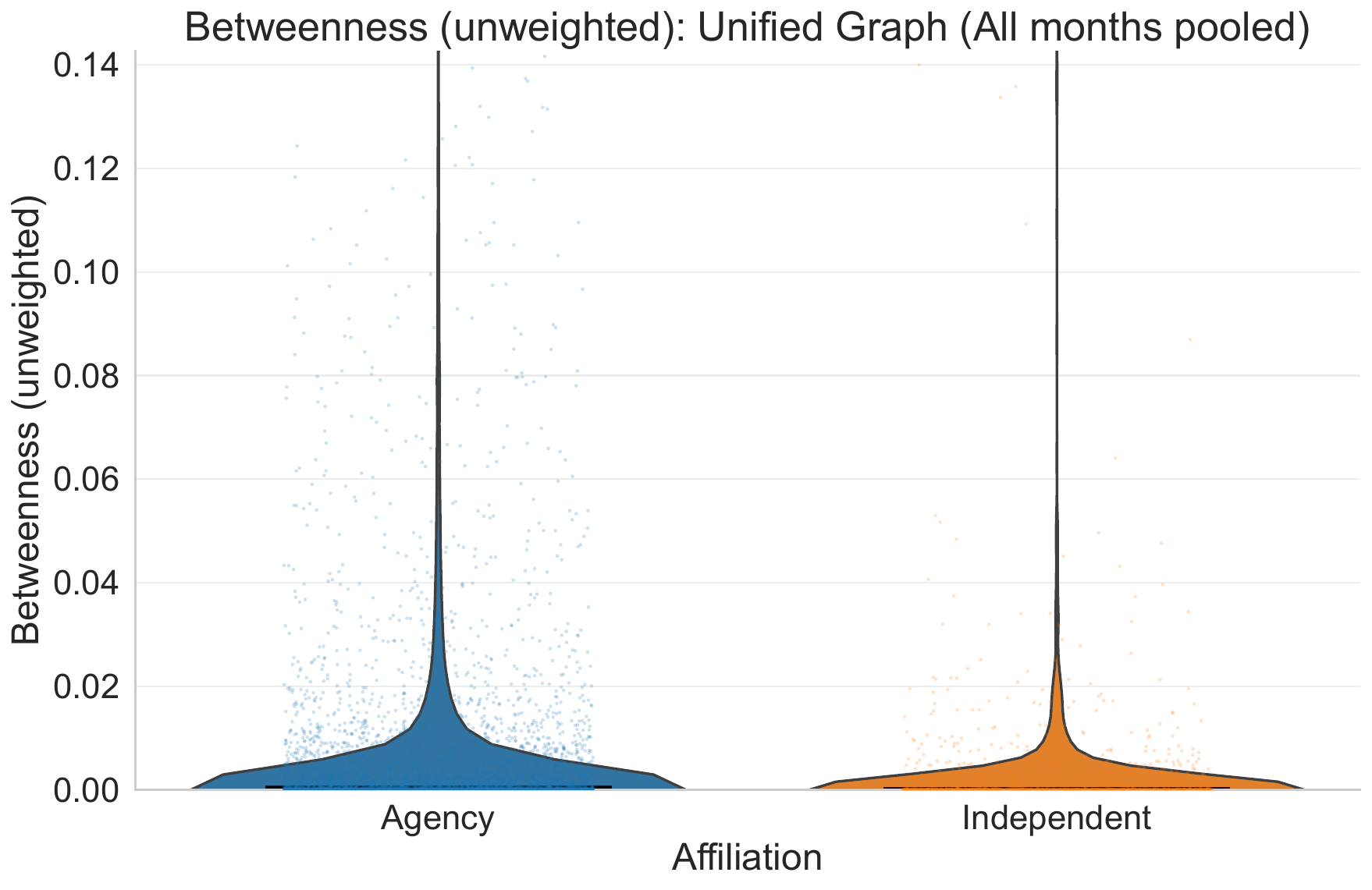}
  }

  \caption{Node-level metrics on the unified audience--overlap network, by affiliation (Agency vs.\ Independent). Each panel shows a violin plot with creator-level points and a horizontal line indicating the mean. Metrics: (a) weighted degree (sum of Simpson weights); (b) degree; (c) local clustering coefficient; (d) unweighted betweenness centrality.}
  \label{fig:unified_node_metrics}
\end{figure}

While these segment-specific subgraphs focus on within-segment cohesion, the unified audience--overlap network clarifies how affiliated and independent creators are positioned in the ecosystem as a whole. \autoref{fig:unified_node_metrics} summarizes node-level measures on the pooled graph. The distributions of weighted degree and degree are shifted upward for agency-affiliated creators, indicating that they share audiences with a larger and more strongly connected set of neighbors. Local clustering coefficients are also higher on average among Agency nodes, consistent with the dense triadic structure visible in the snapshots. By contrast, betweenness centrality is highly skewed for both groups, with most creators lying close to zero and a small number of hubs in each segment accounting for most shortest-path traffic. Taken together, these unified-graph metrics suggest that agencies not only form cohesive portfolios internally but also occupy systematically denser and more tightly clustered positions in the overall audience-overlap network.

\section{Discussion}\label{sec:dis}

\subsection{Synthesis of evidence and mechanism}
Evidence from the multimodal analysis converges on a coherent narrative in which MCN affiliation is associated with a distinct structural and behavioral ecosystem for fans. Structurally, the affiliated segment shows selective local thickening rather than indiscriminate densification. This appears as a decoupling of network metrics, where the average degree in the affiliated subgraph widens relative to the independent subgraph even as their overall edge densities converge (\autoref{fig:network_metrics_timeseries_vertical}; see also \autoref{fig:density}, \autoref{fig:avg_degree}). Network visualizations make the geometry of this trend salient: affiliated creators form a robust core and periphery with well connected hubs and redundant paths, whereas independent creators remain more peripheral and more fragmented (\autoref{fig:vtuber_networks}). This structural signature is mirrored in behavior. Affiliated fans exhibit higher long term retention and, notably, a greater propensity to switch their primary creator (\emph{oshi}) (\autoref{fig:commitment_comparison_grid}). A flow decomposition helps reconcile these findings by showing that mobility is directed inward: movement across segments is limited, and a pattern of ``circulate and recapture'' within the affiliated portfolio appears to mitigate churn (\autoref{fig:sankey_flow_comparison}).

We interpret these convergent patterns as consistent with an organizational infrastructure that cultivates \emph{coordinated proximity}. Through interventions such as cross promotional appearances, synchronized programming, and shared events, MCNs internalize audience spillover and provide fans with nearby alternatives. This portfolio level design can buffer against outright exit by converting the precariousness of individual channels into collective stability. The mechanism offers a plausible account of the observed network dynamics: by thickening ties around curated neighborhoods, MCNs raise local connectedness (average degree) without uniformly increasing the fraction of all possible ties at the segment level (edge density). Enhanced engagement within the affiliated segment is therefore consistent with an emergent property of portfolio level infrastructure that absorbs and redirects fan mobility.

Throughout, we treat “Independent” as a pseudo non-corporate segment—the complement \(A_c=0\)—that aggregates heterogeneous production models (from hobbyists to small studios to “super independents”). Accordingly, between-segment comparisons should be read as \emph{average} gaps across a deliberately coarse divide rather than contrasts between two internally coherent portfolios. This heterogeneity inflates within-segment variance and can attenuate (or, in rare cases, exaggerate) affiliation gaps; interpretations of the \textit{Independent}–\textit{Retain} cell in Fig.~2 should therefore be conservative. Future work will stratify \(A_c=0\) by observable proxies of organizational capacity and re-estimate transition shares and network metrics for robustness.

\subsection{Theoretical and managerial implications}
Theoretically, the findings position MCN affiliation as a critical meso level structure. For labor sociology, the results refine accounts of individualized creator precarity by suggesting that attention risk may be partially pooled: organizational infrastructures redistribute volatility from single channels to portfolios where alternate routes sustain participation \cite{GregersenOrmen2023, Glatt2024}. For marketing and influencer research, affiliation is elevated from a creator attribute to a structural condition that shapes fan mobility, grounded in longitudinal behavioral data rather than cross sectional perceptions \cite{Vrontis2021InfluencerReview, Joshi2025, Kokkodis2023, Sokolova2020, Giertz2022}. For platform studies, the coexistence of local thickening and global convergence, together with recapture dynamics within portfolios, illustrates how intermediaries can work with platform logics to channel algorithmic exposure into denser and more traversable neighborhoods for users \cite{HodlMyrach2023, Yang2022, Koch2018}.

These insights also yield managerial implications. For agencies and MCNs, we suggest strategic investment in \emph{coordinated proximity}: design synchronized schedules, facilitate recurring cross appearances, and organize shared events that shorten distances between hubs within the portfolio \cite{Zhang2024MCN, Koch2018}. Success should be evaluated not only by channel specific KPIs but also by portfolio level health metrics, such as within segment switch rates and mean active run length. Independent creators can approximate these benefits through lightweight alliances, including periodic collaborations or joint programs, in order to build bridges to adjacent clusters and create redundancy against attrition \cite{Koch2018, Giertz2022}. Platforms, in turn, could complement recommendation systems by surfacing portfolio level diagnostics, labeling affiliations transparently, and designing affordances that encourage healthy circulation within portfolios without amplifying market concentration \cite{HodlMyrach2023, Yang2022}.

\subsection{Alternative explanations, limitations, and future work}
While our findings are consistent with an infrastructural account, several alternative mechanisms merit consideration. Selection bias may confound results if more capable creators preferentially join agencies; this concern could be addressed with propensity score matching or with a difference in differences design around affiliation transitions \cite{Kokkodis2023}. Other potential confounders include genre mix, platform wide algorithm shocks, and the intensity of creator supply (for example, total hours streamed), which can be probed via stratification, event study designs, and conditional analyses, respectively \cite{HodlMyrach2023, Wollborn2023, Yang2022}. Our core findings remained robust across a range of network construction thresholds $(\theta, n_{\min}, \tau_U)$, across alternative graph metrics such as clustering coefficients and $k$ core sizes, and across behavioral definitions such as varying retention horizons. These checks reduce concern that the patterns arise from modeling choices.

This study is nonetheless subject to important limitations. First, reliance on chat based traces omits silent viewers and other forms of non chat engagement, which yields an incomplete picture of the audience \cite{Giertz2022}. Second, the \textit{Independent} segment is treated as a single group despite substantial internal heterogeneity, which likely inflates within group variance and may attenuate observed effects. Third, the temporal and cultural scope (YouTube VTubers from 2022 to 2025) constrains generalizability to other platforms, creator types, and periods. Finally, the analyses are descriptive by design and do not claim causal identification.

These limitations suggest a clear agenda for future research. A primary goal is causal identification via natural experiments such as affiliation transitions, large scale collaborations, or platform policy changes \cite{Zhang2024MCN, Koch2018}. The analytical framework could be extended to other platforms (for example, Twitch, Bilibili, or TikTok Live) and to other creative domains (for example, music or live commerce) to test the portability of the ``loose mobility'' mechanism \cite{Wollborn2023}. Future work should also link mobility regimes to creator centric outcomes, including exposure to harassment, burnout, and revenue volatility, in order to assess broader implications for equity and wellbeing \cite{Glatt2024, Takano2024, Ma2022}. A promising translational step is to design and test platform interventions, such as recommendation nudges that preserve diverse paths within portfolios, using randomized controlled experiments to support healthier and more resilient creator ecosystems \cite{HodlMyrach2023, Yang2022}.

\section{Conclusion}\label{sec:con}
This study posed two linked questions and addressed them using a reproducible, fan centered framework. For RQ1, which concerns fan level behaviors, affiliation with agencies is associated with systematically different trajectories: viewers of agency affiliated creators return more often in the short run, with retention higher by about ten to eleven percentage points over two to three months; they switch their primary creator watched (\textit{oshi}) more frequently by roughly 10.7 percentage points; and they exhibit lower inactivity and longer active runs. Taken together, these patterns indicate \emph{loose mobility} rather than lock in to a single channel: movement occurs, but it tends to keep fans in circulation rather than pushing them toward exit.

For RQ2, which concerns meso level structure, monthly audience overlap networks reveal convergence in edge density between the affiliated and independent segments, while local connectivity remains markedly different. The affiliated subgraph sustains a substantially higher mean degree, on the order of $6.8\times$ that of independents, with a persistent gap in clustering as well. Segment level state flow diagrams align with these structural differences by showing reduced cross segment leakage and fewer transitions into inactivity where agencies operate. Taken together, the structural and behavioral results are consistent with proximity engineered by agencies that facilitates movement within portfolios and \emph{recaptures} attention.

The joint answers to RQ1 and RQ2 therefore support a coherent account: agencies function as organizational infrastructures that pool attention risk by thickening local neighborhoods among portfolio creators, enabling fans to move in order to stay. We do not claim causality. Our contribution is a reusable measurement framework that links micro level trajectories to meso level structure and documents robust empirical regularities on which future identification strategies, including designs around affiliation changes, natural experiments, or platform interventions, can build. These findings speak to debates on creator precarity, influencer marketing, and platform governance by showing how portfolio organization shapes both the circulation of audiences and the topology that channels their movement.

\bibliographystyle{IEEEtran}
\bibliography{main}

\vspace{12pt}
\end{document}